\documentclass[fleqn,usenatbib]{mnras}

\usepackage{newtxtext,newtxmath}

\usepackage[T1]{fontenc}

\DeclareRobustCommand{\VAN}[3]{#2}
\let\VANthebibliography\thebibliography
\def\thebibliography{\DeclareRobustCommand{\VAN}[3]{##3}\VANthebibliography}

\usepackage{graphicx}	
\usepackage{amsmath}	
\usepackage{orcidlink}  
\usepackage{multirow}   
\usepackage{textgreek}  
\usepackage[export]{adjustbox} 

\newcommand{\lnL}{\textrm{ln}\mathcal{L}} 

\title[Gaia DR3 Brown Dwarfs]{Combing the Brown Dwarf Desert with \emph{Gaia} DR3}

\author[A. T. Stevenson et al.]{
Adam T. Stevenson$^{1}$\thanks{E-mail: adam.stevenson@open.ac.uk}\orcidlink{0000-0003-2399-7619},
Carole A. Haswell$^{1}$\orcidlink{0000-0002-8050-1897},
John R. Barnes$^{1}$\orcidlink{0000-0001-6105-2902} and
Joanna K. Barstow$^{1}$\orcidlink{0000-0003-3726-5419}
\\
$^{1}$School of Physical Sciences, The Open University, Milton Keynes MK7 6AA, UK
}

\date{Accepted XXX. Received YYY; in original form ZZZ}

\pubyear{2023}

\begin{document}
\label{firstpage}
\pagerange{\pageref{firstpage}--\pageref{lastpage}}
\maketitle

\begin{abstract} 
We have created an up-to-date catalogue of 214 brown dwarfs (BDs) in binaries with $P < 10^{4}$\,d. This allows us to examine the population statistics of the brown dwarf desert. We searched \textit{Gaia} DR3 NSS results for orbital inclinations of BD candidates, deriving 12 new masses. Three remain as desert BDs whereas nine candidates are found to be low-mass stars. 
We improved the RV solutions for three previously studied BD candidates. A further 19 BD masses with periods less than $\sim 1200$\,d were identified in the DR3 \texttt{binary\_masses} database.
We confirm a valley in the mass distribution with a minimum around $30$--$35$~M$_{\textrm{jup}}$, and find that periods $<100$~d are still under-populated in comparison to longer periods. The updated mass and eccentricity distribution of BDs still shows a marginally statistically significant split into high- and low-mass BD populations. This hints at two different parent distributions, and two potential origins -- either akin to planetary formation, or stellar. There are 
no low eccentricity BDs at periods around $100$~d.
The mass-metallicity distribution of BDs indicates that core accretion is not the dominant formation mechanism for BDs as they do not follow the same trends that giant exoplanets do with metallicity.
We identify a diagonal envelope bounding the \textit{Gaia} BDs in the mass-period plane due to the detection thresholds of the currently available NSS solutions from 34 months of data.
\end{abstract}




\section{Introduction}\label{sec:introduction}
Brown dwarfs (BDs) are celestial objects with masses lying between those of planets and stars. The upper bound of the BD mass range is the hydrogen-burning minimum mass (HBMM), the mass required to ignite and sustain stable nuclear fusion of hydrogen to helium in the core of a star \citep{Kumar1963,HN1963,CB1997}. BDs lack the mass and self gravity to contract to a sufficient density (and high enough core temperature) before electron degeneracy pressure halts the collapse. Objects below $\sim80$~M$_{\textrm{jup}}$ are unable to fuse hydrogen, and instead can only fuse deuterium (and lithium if massive enough) in their core. The minimum mass and core temperature required to fuse deuterium (henceforth DBMM) defines the lower bound of the BD mass range, separating them from the most massive giant planets. This mass depends on a number of factors, including helium and deuterium abundances, and metallicity of the individual objects, but is usually taken as $\sim13$~M$_{\textrm{jup}}$ \citep{Burrows2001,Spiegel2011}. This limit however does not have strong physical justification \citep{2004ASPC..321..131C,Wilson2016}, and the planet-BD boundary would be better defined through formation mechanism. Studying objects with masses in this region is therefore critically important for theories of sub-stellar objects.

Radial velocity (RV) studies reached the required precision to detect sub-stellar companions through host reflex orbital motions decades ago, revealing a notable deficit of BDs in close orbits around host stars \citep{Campbell1988}. The paucity was coined as a `brown dwarf desert', and is usually defined to exist up to orbital semi-major axes of $\sim5$~au \citep{Wilson2016}. BD companions on longer orbits were seen to be far more common, as were planetary and stellar companions across the full range of orbital periods \citep{GretherLineweaver2006,Feng2022}. This deficit seems to have no conclusive explanation so far \citep{Grieves2017}, 
but could indicate the presence of two distinct mass distributions, for brown dwarfs that form like planets and those that form like stars.

As the required masses for deuterium and hydrogen burning only come into play \textit{after} the formation of the object in question, there is no reason that the formation itself should be impacted by the fairly arbitrary definition of a `brown dwarf'. \citet{GretherLineweaver2006} analysed the BD desert and found that the companion mass functions for planets and stars both decreased towards the BD mass range. The sum of the two mass distributions produced a minimum number of companions per unit interval in log mass at $\sim31$~M$_{\textrm{jup}}$. The frequency of BD companions around sun-like stars was observed to be $\lesssim 1$~per~cent, compared with the $16$~per~cent of stars having any type of companion at the time. It is currently accepted that 40--50 per cent of sun-like stars have a stellar companion \citep{DK2013, Matson2018}, and 35--75 percent of stars studied by \textit{Kepler} have `Kepler-like' planets \citep{Yang2020}. Despite such a prevalence of companions, the occurrence of circumstellar BDs remains close to $1$~per~cent, in either large RV or imaging surveys \citep{Nielsen2019,Feng2022}.

The BD desert was originally considered an observational bias, but this explanation is now less plausible. RV detections are most sensitive to short period, high mass objects, so should detect BD companions more easily than planets were the desert not real\footnote{Though some stars with large RV excursions indicating high mass companions may have been abandoned in RV planet searches without being published; DMPP-3 provides an example of this \citep{Barnes20}.} \citep{MaGe2014}. Numerous BDs are observed free-floating throughout our Galaxy, not orbiting a host star. The occurrence distribution of field stars/BDs does not tail off towards low-mass objects in the same manner it does for stellar companions \citep{GretherLineweaver2006,Feng2022}, perhaps indicating different formation mechanisms between close-orbiting circumstellar BDs and those formed in fields and clusters \citep{MaGe2014}. 

Determining the distribution function of brown dwarfs in the period-mass parameter space requires a large sample of objects, and is vital to study the origins of the brown dwarf desert \citep{Feng2022}. To do this, we require constrained masses of objects, through combination with inclination. Minimum masses from RV studies are a factor of $\sin{i}^{-1}$ less than the mass. Many provisional members of the brown dwarf desert could in-fact be stellar companions in nearly face-on orbits, with minimum masses in the BD mass range \citep{Halbwachs2000}. Any stars masquerading as BD candidates, which only have minimum mass information, will hopelessly blur the BD desert demographics. We therefore need a sample of accurate BD masses, which we expand upon in this work.

The orbital inclination angle can be determined if we observe a BD transiting its host star, or by precisely measuring the positions and motions with astrometry. With the third data release (DR3; \citealt{DR3paper}) from \textit{Gaia} and the associated binary star processing, we can deduce mass information from the orbital analysis of positions and motions, e.g. \citet{Halbwachs2023,Xiao2023}. DR3 binary results have enabled inclinations to be calculated down to very low mass stars (VLMS) and beyond into the brown dwarf regime. We recently published a paper in which the determination of $i$ from \textit{Gaia} DR3 allowed us to refine the mass of DMPP-3\,B, an object which previously lay on the HBMM threshold \citep{Stevenson2023}. This prompted us to examine whether there were other putative BDs companions whose masses we could determine, and we apply the same analysis to objects in DR3 hosting candidates with minimum masses in or close to the BD desert.

In Section~\ref{sec:method} we outline our use of \textit{Gaia} data products and our method for calculating masses of BD candidates. In Section~\ref{sec:catalogue} we describe the catalogue of BDs/BD candidates created to search for constrained masses and use in demographic studies of the desert. In Sections~\ref{sec:BDverification} and \ref{sec:BD_bin_mass} we determine masses for previously detected and new BD candidates, respectively. The analysis of the BD desert is discussed in Section~\ref{sec:discussion}, with limitations to this work discussed in Section~\ref{sec:limitations}. We conclude and summarise our findings in Section~\ref{sec:conclusions}.

\section{Gaia}\label{sec:method}
The \textit{Gaia} mission was launched near the end of 2013 to create the largest map of the Milky Way to date \citep{GaiaMission}. The goal was to determine the 3D spatial and 3D velocity distribution of stars for a representative part of the Galaxy. To do this, high-precision astrometry was required -- the accurate measurement of positions, and how they change with time. This is not possible from ground based observatories, due to the varying nature of Earth's atmosphere, so the \textit{Gaia} space observatory resides at the Sun-Earth L2 Lagrange point in a Lissajous-type orbit.

The most recent data release (DR) from \textit{Gaia}, DR3, contains results derived from 34 months of satellite operations, and includes astrometry for more than $1.8\times10^{9}$ sources \citep{DR3paper}. For bright enough stars, the observations are complemented by radial velocity measurements, resulting in 882 million sources with 6-parameter astrometry. The datasets created from DR3 are vast, and can be used to learn about the formation, structure, and evolution of the Galaxy \citep{GaiaMission}. Sub-sets of stars have been processed in various ways, producing astrophysical parameters \citep{Creevey2023}, searching for variable sources \citep{Rimoldini2023}, or as used herein, finding stellar companions \citep{Halbwachs2023,GaiaStellarMulti2023}.

\subsection{NSS}
The non-single star (NSS) processing performed for DR3 is described in works such as \citet{Halbwachs2023} and \citet{GaiaStellarMulti2023}. The end result is a catalogue of $\sim813000$ sources with astrometric orbits,  single or double-lined spectroscopic orbits (SB1 and SB2 respectively), eclipsing binary solutions, or acceleration trends \citep{DR3paper}. To be considered for NSS analysis, the motion must significantly deviate from single star models. This is quantified through use of a goodness-of-fit metric \citep{Almenara2022}, known as the re-normalised unit weight error (RUWE). For RUWE $\gtrsim 1.4$ there is sufficient excess noise to potentially be caused by an orbiting body \citep{Lindegren2021}. If the star passes filtering and significance checks \citep{GaiaStellarMulti2023}, it can then be processed to determine the companion orbit.

For stars with fitted astrometric orbits, the parameters of such an orbit in the NSS databases\footnote{\texttt{NssTwoBodyOrbit\_1} downloaded from \url{http://cdn.gea.esac.esa.int/Gaia/gdr3/Non-single_stars/nss_two_body_orbit/}} are expressed in Thiele-Innes elements ($A,B,F,G,C,$ and $H$). The appendices of \citet{Halbwachs2023} describe the conversion of these to traditional Campbell orbital elements ($a,\omega,\Omega,$ and $i$), along with respective measurement uncertainties. This can also done with the python package \textsc{nsstools}\footnote{\url{https://gitlab.obspm.fr/gaia/nsstools}} \citep{Halbwachs2023}. These parameters, alongside the NSS periods and eccentricities, allows us to compare solutions with RV-detected BD candidates throughout literature, and determine masses for some of these objects.

\subsection{Radial velocity masses}

The RV observable, semi-amplitude ($K$), is proportional to $M_{2}\sin{i}$, where $M_{2}$ is the mass of the companion and $i$ is the inclination of the orbit \citep{Perryman2000}. These are considered as minimum masses, and require further information to determine the object's mass. Throughout this paper, unless we refer to minimum mass, when we mention mass it refers to that constrained by combining inclination with RV observations, or determined with astrometric measurements.

As secondary stars/BDs have non-negligible mass with respect to the host, we require a numerical method to solve for $M_{\textrm{2}}$. In this paper we have used the Newton-Raphson process \citep{Dedieu2015,2020SciPy-NMeth} to solve the functional form $f$, shown here as equation~\ref{eqn:N-R}.

\begin{equation}\label{eqn:N-R}
    f = \frac{M_{\textrm{2}}^3\,\sin^3 i}{(M_{\text{1}}+M_{\text{2}})^2} = \frac{P_{\text{orb}}K^3}{2 \pi G}\sqrt{1-e^{2}}.
\end{equation}

\noindent This method has been employed where we have had to re-calculate minimum masses from archival RV observations in this paper. 
The posterior distributions of the orbital solutions are sampled by Markov chain Monte Carlo (MCMC) techniques \citep{emcee}, and corresponding error analysis is performed by calculating masses for each sample in the MCMC chains. For the posterior parameter exploration performed, we used fairly wide uniform priors informed by solutions from the NSS database. Using the analysis in Section~\ref{subsub:HD105963} as an example, we chose $P\in [0,10000]$~d, $K\in [0,20000]$~m s$^{-1}$, $e\in[0,0.5]$, $\omega\in[0,2\pi]$, $M_{\textrm{0}}\in[0,2\pi]$.

\subsection{Companion Masses}\label{sub:true-masses}

Measurements of orbital inclination allow us to solve for the mass of a companion, given that the $M\sin i$ is known. Here and hereafter we are dropping the subscript on $M_{2}$, but we are referring throughout to the masses of secondary components orbiting stars with {\it Gaia} astrometry. The error calculation is more involved, and for this we have chosen to use symmetrical errors in $M\sin i$, taking the largest value for simplicity when two different values were quoted in literature. Inclination errors output from \textsc{nsstools} are already symmetrical. Errors on mass have been calculated via equation~\ref{eqn:dM}:

\begin{equation}\label{eqn:dM}
    \Delta M = M \sqrt{\frac{(\Delta~M\sin i)^2}{(M\sin i)^2} + \frac{(\cos i)^2 (\Delta i)^2}{(\sin i)^2}}.
\end{equation}

As a secondary evaluation of masses, we can use the \texttt{binary\_masses}\footnote{\texttt{BinaryMasses-001} table downloaded from \url{http://cdn.gea.esac.esa.int/Gaia/gdr3/Performance_verification/binary_masses/}} table provided in DR3, as a supplement to the NSS. Many stars identified as binaries are listed in the mass table, with either $M_{1}$ \& $M_{2}$, or $M_{1}$ and limits on $M_{2}$. The primary masses are predominantly derived in the DR3 pipeline through isochrone fitting. This is discussed in Appendix~E of \citet{GaiaStellarMulti2023}, and we direct the reader to that work for a full description. Briefly, \textsc{parsec} isochrones \citep{PARSEC} are combined with the observables of absolute magnitude ($M_{\textrm{G}}$) and colour $(BP-RP)_{0}$ to evaluate the star's place on the H-R diagram. After excluding stars off the main-sequence, the mass estimates are computed from evolutionary models and the physical observables. For special cases where there is an orbital+SB2 solution, or eclipsing SB2 solution, the primary mass can instead be derived directly from the NSS information.

\section{The brown dwarf companion catalogue}\label{sec:catalogue}
We have scoured recent literature in an attempt to collate as large a sample of BD companions as possible. The selection criteria has been for masses/minimum masses between 13--80~M$_{\textrm{jup}}$ \cite[as in][]{GretherLineweaver2006}, and periods of less than $10^{4}$ days. The large period limit was chosen to catch all BDs from other works, where the extent of the brown dwarf desert is not always consistent. For specific uses the sample can be restricted as required.

\begin{figure}
    \centering
    \includegraphics[width=0.99\columnwidth]{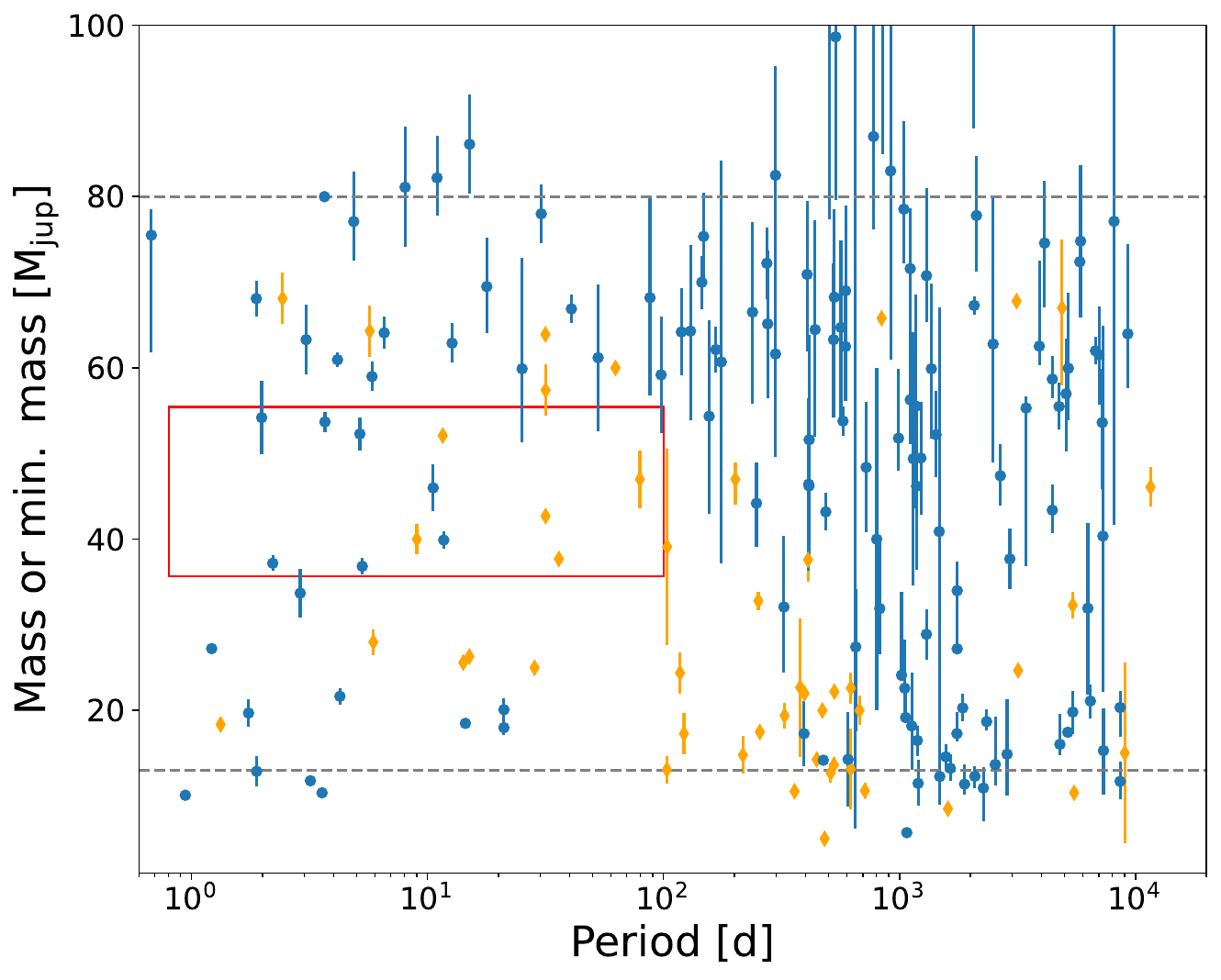}
    \caption{Period-mass diagram for all BDs collated in our sample. 162 blue circles represent BDs with measured mass (including those constrained in this study), whereas the 53 orange diamonds represent those with only minimum mass ($M\sin{i}$). The red box highlights the `driest' region of the desert identified in \citet{MaGe2014}, $0 \leq P \leq 100$~d \& $35 \leq M \leq 55$~M$_{\textrm{jup}}$.}
    \label{fig:MvsP}
\end{figure}

Previous studies of the BD desert have typically still included minimum masses for BDs detected solely through RV studies. To assess how many companion orbits the \textit{Gaia} NSS has automatically characterised, our sample included these RV only detections. However, unless they have DR3 inclinations they will be omitted for investigating population statistics of the desert in Section~\ref{sec:discussion}. BD candidates with no constrained mass from either literature or this work are shown alongside confirmed BDs in Fig.~\ref{fig:MvsP}. The sample has been updated with information from the DR3 databases (see Sections~\ref{sec:BDverification} and~\ref{sec:BD_bin_mass}), and includes a total of 215 companions that fall into one of three groups: candidate BDs; confirmed BDs; or former BD candidates promoted to the stellar regime. The catalogue can be accessed in machine readable format from \url{https://github.com/adam-stevenson/brown-dwarf-desert/tree/main}.

\subsection{Lensing BDs}
In this work, we have chosen not to include those BDs discovered through gravitational microlensing. Despite masses being estimated, these companions have no direct period or eccentricity measurement, so cannot be compared in the same manner as the rest of the sample. Lensing BDs, and their estimated periods from projected separations, are described in other works that we instead direct the reader to (e.g. \citealt{Ranc2015,Ryu2017,Han2022,Herald2022}).

\subsection{NSS Stars}
Of the 195 previously known BDs/BD candidates in the catalogue, 32 are found to be included in the NSS database with sufficient astrometric measurements to constrain the inclination ($\sim 18$~per~cent of objects). 20 of these have masses constrained in other works, either using the NSS results directly, or via proper motion/acceleration fitting \citep{Halbwachs2023, Xiao2023}. The remaining 12 stars and companions will be studied in this paper, where host properties are described in Table~\ref{tab:nssStarParams}. In-depth consideration of individual companions is addressed in Sections~\ref{sub:BDconf} and \ref{sub:starconf}.

\renewcommand{\arraystretch}{1.5}
\begin{table*}
	\centering
	\caption{Host star properties for the companions discussed in Section~\ref{sec:BDverification}. (1) \citealt{Wilson2016}, (2) \citealt{Latham2002}, (3) \citealt{Nidever2002}, (4) \citealt{MaGe2014}, (5) \citealt{Jiang2013}, (6) \citealt{Grandjean2021}, (7) \citealt{Jenkins2009}, (8) \citealt{Grieves2017}. SIMBAD data for V-band and G-band magnitudes, as well as spectral type (SpT), are accessed from~\url{http://simbad.u-strasbg.fr}.}
	\label{tab:nssStarParams}
	\begin{tabular}{lcccccccc} 
		\hline
		Host Star & Mass (M$_{\sun}$) & $T_{\textrm{eff}}$ (K) & log\,$g$ (cm s$^{-2}$) & [Fe/H] & $V_{\textrm{mag}}$ & $G_{\textrm{mag}}$ & SpT &  Ref. \\
		\hline
		  HD\,39392  & $1.08 \pm 0.08$ & $5951 \pm 42$ & $4.08 \pm 0.03$ & $-0.32 \pm 0.03$ & $8.38$ & $8.26$ & F8 & (1)   \\
		HD\,132032 & $1.12 \pm 0.08$ & $6035 \pm 33$ & $4.45 \pm 0.05$ & $0.22 \pm 0.03$ & $8.11$ & $7.97$ & G5 & (1,2)   \\
		HD\,140913 & $0.98$ & $6048 \pm 100$  & $4.57 \pm 0.1$ & $0.07 \pm 0.1$ & $8.053$ & $7.92$ & G0V & (3,4) \\
		HIP\,67526 & $1.11\pm 0.08$ & $6004 \pm 29$  & $4.55 \pm 0.15$ & $0.04 \pm 0.05$ & $9.71$ & $9.58$ & G0V & (4,5)  \\
        BD+24\,4697 & $0.75 \pm 0.05$ & $5077 \pm 32$ & $4.67 \pm 0.07$ & $-0.11 \pm 0.02$ & / & $9.49$ & K2 & (1) \\
        BD+26\,1888 & $0.76 \pm 0.08$ & $4748 \pm 87$ & $4.30 \pm 0.20$ & $0.02 \pm 0.04$ & / & $9.48$ & K7 & (1) \\
        HD\,30339  & $1.1$ & $6074 \pm 100$ & $4.37 \pm 0.1$ & $0.21 \pm 0.1$ & $8.21$ & $8.07$ & F8 & (3,4) \\
        HD\,105963\,A & $0.75$ & / & / & / & / & $7.73$ & K0V/K2 & (6) \\
        HD\,160508 & $1.25 \pm 0.08$ & $6212 \pm 30$ & $4.16 \pm 0.03$ & $0.01 \pm 0.02$ & 8.11 & $7.97$ & F8V & (1)  \\
        HD\,191760 & $1.28^{+0.02}_{0.10}$ & $5794 \pm 76$ & $4.13^{+0.05}_{-0.04}$ & $0.29 \pm 0.07$ & $8.25$ & $8.12$ & G3IV/V  &  (7) \\
        TYC~0173-02410-1 & $1.06 \pm 0.1$  & $5470 \pm 112$ & $4.25 \pm 0.18$ & $0.18 \pm 0.09$ & $10.38$ & $10.23$ & G8V & (8) \\
        GSC~03467-00030 & $0.96 \pm 0.09$  & $5525 \pm 71$ & $4.39 \pm 0.23$ & $-0.17 \pm 0.08$ & $12.07$ & $11.94$ & G6V & (8) \\
		\hline
	\end{tabular}
\end{table*}

\section{Brown Dwarf Verification}\label{sec:BDverification}
Other recent studies have assessed the masses of companions, finding planets that become BDs, and BDs that become stars. \citet{Feng2022} and \citet{Xiao2023} use the astrometric measurements from both \textit{Hipparcos} and \textit{Gaia} to derive these masses, but limit their work to periods longer than $1000$~d. To truly investigate the BD desert, we need to extend mass analysis further down to short periods, particularly $<100$~d, where \citet{MaGe2014} identified the driest region for the mass (including minimum masses) range $35 \leq M \leq 55$~M$_{\textrm{jup}}$. We therefore use the NSS results to determine the mass, and hence nature, of objects not included in these previous works. We have not performed joint fits of RV and astrometry measurements in our analysis, as this is beyond the scope of this work. The individual \textit{Gaia} epoch RV and positions are also not published in their entirety until the release of DR4.

We have determined masses for 12 companions where these had not been previously determined. Only three of these 12 objects remain as a BD. The orbital parameters and masses of these companions are described in the following sub-sections, and are fully tabulated in Table~\ref{tab:NewTrueMasses}. That 75~per~cent of these candidates are in fact low-mass stars highlights the point that only well-constrained masses should be used to assess the desert. The desert could be more arid than previously thought.

\subsection{New masses: BD confirmation}\label{sub:BDconf}

\renewcommand{\arraystretch}{2}
\setlength{\tabcolsep}{3pt}
\begin{table*}
	\centering
	\caption{Tabulated orbital parameters and masses for the companions discussed in Section~\ref{sec:BDverification}. For each star, companion solution references are the same as in Table~\ref{tab:nssStarParams}. (*): literature parameters not listed for HD105963A, and have been calculated from archival RVs in this work. They are tabulated here to compare an RV solution to the \textit{Gaia} orbital solution. (**): Re-derived in this work from archival RVs, to be consistent with orbital solution found by \textit{Gaia}.}
	\label{tab:NewTrueMasses}
	\hspace*{-0.6cm}\begin{tabular}{lcccccccccc}
		\hline
		Host Star & $P_{\textrm{lit}}$~(d) & $P_{\textrm{Gaia}}$~(d)  & $e_{\textrm{lit}}$ & $e_{\textrm{Gaia}}$ & $M\sin i$~(M$_{\textrm{jup}}$) & $i~(^\circ)$ & $M$~(M$_{\textrm{jup}}$)\\
		\hline
        HD\,39392 & $394.3_{-1.2}^{+1.4}$ & $389.19 \pm 2.03$  & $0.394 \pm 0.008$ & $0.153 \pm 0.241$ & $13.2 \pm 0.8$ & $49.72 \pm 14.41$ & $17.30 \pm 3.84$ \\
		  HD\,132032 & $274.33 \pm 0.24$  & $274.41 \pm 2.51$  & $0.0844 \pm 0.0024$ & $0.090 \pm 0.087$ & $70 \pm 4$ & $75.73 \pm 2.18$ & $72.23 \pm 4.19$ \\
		HD\,140913 & $147.968$ & $147.632 \pm 0.151$  & $0.54$ & $0.513 \pm 0.046$ & $43.2$ & $34.98 \pm 2.74$ & $75.36 \pm 5.15$\\
        HIP\,67526 & $90.2695 \pm 0.0188$ & $90.226 \pm 0.159$  & $0.4375 \pm 0.004$ & $0.384 \pm 0.052$ & $62.6 \pm 0.6$ & $11.49 \pm 12.82$ & $314.21_{-163.61}^{+981}$\\
		BD+24\,4697 & $145.081 \pm 0.016$ & $145.24 \pm 0.101$  & $0.5005 \pm 0.0004$ & $0.483 \pm 0.014$ & $53 \pm 3$ & $161.53 \pm 2.25$ & $167.31 \pm 21.86$\\
        BD+26\,1888 & $536.78 \pm 0.25$ & $537.31 \pm 1.07$  & $0.268 \pm 0.002$ & $0.275 \pm 0.014$ & $26 \pm 2$ & $15.27 \pm 2.76$ & $98.66 \pm 19.00$\\
        HD\,30339 & $15.0778$ & $15.1039 \pm 0.0151$  & $0.25$ & $0.165 \pm 0.159$ & $77.8$ & $64.61 \pm 8.08$ & $86.12 \pm 5.76$\\
        HD\,105963\,A & $642.10 \pm 0.97$* & $646.01 \pm 0.62$ & $0.028 \pm 0.014$* & $0.010 \pm 0.003$ & $567.7_{-22.8}^{+24.1}$* & $125.17\pm 0.11$ & $664.4 \pm 29.5$\\
        HD\,160508 & $178.905 \pm 0.007$ & $178.309 \pm 0.278$  & $0.5967 \pm 0.0009$ & $0.575 \pm 0.048$ & $48\pm 3$ & $169.83 \pm 10.15$ & $271.81^{+268.95}_{-142.23}$\\
        HD\,191760 & $505.65 \pm 0.42$ & $506.38 \pm 3.15$  & $0.63 \pm 0.01$ & $0.625 \pm 0.044$ & $38.17\pm 1.02$ & $158.96 \pm 5.99$ & $106.34 \pm 29.03$\\
        TYC~0173-02410-1 & $216.5 \pm 1.8$ & $249.71 \pm 0.33$  & $0.37\pm0.05$ & $0.337 \pm 0.019$ & $81.75 \pm 7.75$** & $13.02 \pm 6.05$ & $362.86 \pm 169.25$\\
        GSC~03467-00030 & $147.6 \pm 0.3$ & $295.77 \pm 0.56$  & $0.50 \pm 0.04$ & $0.709 \pm 0.022$ & $53.59^{+7.33}_{-5.50}$** & $159.17 \pm 3.38$ & $150.68 \pm 31.16$\\
		\hline
	\end{tabular}
\end{table*}

\subsubsection{HD\,39392}
The BD companion to HD\,39392 is listed in \citet{Wilson2016} as having $M\sin{i} = 13.2 \pm 0.8$~M$_{\textrm{jup}}$. Orbital parameters are given as $P = 394.3^{+1.4}_{-1.2}$~d, and $e = 0.394 \pm 0.008$. \textit{Gaia} analysis reports a very similar period of $389.19$~d, but cannot constrain the eccentricity well, with $e = 0.15 \pm 0.24$. The inclination is calculated to be $i = 49.72 \pm 14.41^{\circ}$, with large errors likely a result of a poor fit to the orbital eccentricity. This could stem from the small astrometric wobble for a low-mass BD. The inclination allows us to solve for mass, $M = 17.30 \pm 3.84$~M$_{\textrm{jup}}$. The object is therefore confirmed as a BD, as the lower bound on mass still places it above the DBMM threshold.

\subsubsection{HD\,132032}\label{subsub:HD132032}
HD\,132032 hosts a BD candidate that lies near the hydrogen-burning threshold \citep{Latham2002,Wilson2016}. Analysis of the orbit with \textit{Gaia} NSS supports the previously determined parameters, having excellent agreement in $P$ and $e$ (Table~\ref{tab:NewTrueMasses}). The mass of $M\sin{i} = 70 \pm 4$~M$_{\textrm{jup}}$ is only increased slightly with inclination ($i = 75.73 \pm 2.18^{\circ}$), resulting in a mass of $M = 72.23 \pm 4.19$~M$_{\textrm{jup}}$. This supports the BD classification, but is on the fringes of the HBMM. The \texttt{binary\_masses} table provides a companion mass value of $M = 70.38_{-4.21}^{+4.18}$~M$_{\textrm{jup}}$, in good agreement with the combined RV+inclination value.

\subsubsection{HD\,140913}
The companion to HD\,140913 was reported by \citet{Nidever2002} and \citet{MaGe2014}. It was placed squarely in the middle of the BD mass regime ($M\sin{i} = 43.2$~M$_{\textrm{jup}}$). With $P \sim 148$~d (in agreement between previous and current work) this object would be just outside the `driest' region of the BD desert in mass-period space. The mass was very close to the \citet{MaGe2014} transition between formation mechanisms (see Section~\ref{sec:discussion}) and it has period only 1.5 times longer than the fairly arbitrary cutoff at $P = 100$~d (cf. Fig.~\ref{fig:MvsP}).

Combined RV and \textit{Hipparcos} analysis by \citet{Sahlmann2011} has already provided an upper limit to the mass of this companion as $0.56~$M$_{\sun}$ ($586.6$~M$_{\textrm{jup}}$), given the weak significance of their orbital fit. We find the mass, given inclination from NSS of $i = 34.98 \pm 2.74^{\circ}$, to be $M = 75.36 \pm 5.15$~M$_{\textrm{jup}}$.

The \texttt{binary\_masses} table estimate is $M = 86.25_{-7.51}^{+7.13}$~M$_{\textrm{jup}}$. Therefore, depending on what mass estimate we use, the companion to HD\,140913 is either just under the HBMM (and a BD), or just over it (and a VLMS). The discrepancy is likely to arise from the primary mass of HD\,140913 itself. \citet{MaGe2014} report a primary mass of $0.98~$M$_{\sun}$, whereas in the DR3 database (isochrone fitting) it is listed as $1.05~$M$_{\sun}$.

\subsection{New masses: BD candidates shown to be stars}\label{sub:starconf}

\subsubsection{HIP\,67526}
The companion to HIP\,67526 \citep[MARVELS-5:][]{Jiang2013,MaGe2014} is a brown dwarf candidate with minimum mass $M\sin{i} = 62.6 \pm 0.6$~M$_{\textrm{jup}}$. Period and eccentricity are consistent between previous studies and the \textit{Gaia} NSS results (Table~\ref{tab:NewTrueMasses}). 

The inclination is found to be $i = 11.49 \pm 12.82^{\circ}$, therefore mass of this object is estimated to be $M = 314.21$~M$_{\textrm{jup}}$. 
Due to a nearly face-on inclination and large uncertainty in $i$, equation~\ref{eqn:dM} gives a mass uncertainty greater than mass value itself, $\Delta M = 345.89$~M$_{\textrm{jup}}$. To avoid negative masses, and because we have a definite minimum mass from RVs, we determine the lower bound on mass in an alternative manner. We instead combine the lower bound of $M\sin{i}$ ($62$~M$_{\textrm{jup}}$) with the least face-on inclination ($i\sim 24.31^{\circ}$). This results in a lower mass bound of $150.60$~M$_{\textrm{jup}}$.

Additionally, the upper bound on mass will be functionally infinite as this error on orbital inclination includes configurations that are face-on (and beyond). We instead use the wide bounds provided in the \textit{Gaia} \texttt{binary\_masses}. This is tabulated as [137,1295]~M$_{\textrm{jup}}$. Astrometric minimum mass is relatively close to the RV-derived minimum (and crucially still over the HBMM threshold), and the maximum provides a weak but nonetheless useful constraint on the largest possible mass. Despite the large uncertainty bounds, this object can be reclassified from BD to low-mass star.

\subsubsection{BD+24\,4697}\label{subsub:BD+244697}
\citet{Wilson2016} reported a companion to BD+24\,4697 (HIP\,113698) on a $P = 145$~d, $e = 0.5$ orbit. With a host mass of $M_{\textrm{star}} = 0.75 \pm 0.05$~M$_{\sun}$, they derived a minimum mass for the brown dwarf candidate of $M\sin{i} = 53 \pm 3$~M$_{\textrm{jup}}$. \textit{Hipparcos} measurements provide a loose upper limit of $0.51$~M$_{\sun}$ ($534$~M$_{\textrm{jup}}$) for the companion \citep{Wilson2016}. 

\textit{Gaia} data confirms the orbital parameters $P$ and $e$, and measures inclination for the system ($i = 161.53 \pm 2.25^{\circ}$). The resulting mass for the companion is therefore $M = 167.31 \pm 21.86~$~M$_{\textrm{jup}}$. This is in good agreement with masses from isochrone fitting and astrometry, as the \texttt{binary\_masses} table provides a value of $M = 172.29^{+29.97}_{-28.72}$~M$_{\textrm{jup}}$.

\subsubsection{BD+26\,1888}
A companion to the star BD+26\,1888 (HIP\,44387) was also reported by \citet{Wilson2016} as a brown dwarf candidate. The minimum mass from radial velocities was found to be $M\sin{i} = 26 \pm 2$~M$_{\textrm{jup}}$. The period and eccentricity were $537$~d and $0.27$, respectively. 

The orbital parameters are in excellent agreement with those from \textit{Gaia} DR3, and the measured inclination of $i = 15.27 \pm 2.76^{\circ}$ gives a mass of $M = 98.66 \pm 19.00$~M$_{\textrm{jup}}$. The lower limit lies just at the HBMM, with large errors from the uncertainty on the highly inclined orbit. 

Previous \textit{Hipparcos} measurements constrained the upper limit of the companion mass to be $0.25$~M$_{\sun} (262$~M$_{\textrm{jup}}$) \citep{Wilson2016}. \textit{Gaia} mass derivations (\texttt{binary\_masses}) give limits on companion mass as [$135, 1096$]~M$_{\textrm{jup}}$. This is not strictly in agreement with masses from radial velocity, solved with astrometrically calculated inclination. As the \textit{Gaia} masses are larger, we assume that the lower limit of $M = 79.66$~M$_{\textrm{jup}}$ is likely a conservative estimate, and the object will lie comfortably over the hydrogen burning threshold -- and therefore can be re-classified as a low-mass stellar object.

\subsubsection{HD\,30339}
The companion to HD\,30339 was reported as a brown dwarf object, with mass straddling the hydrogen-burning threshold, $M\sin{i} = 77.8$~M$_{\textrm{jup}}$ \citep{Nidever2002,Wilson2016}. Joint RV and astrometric solutions were attempted by \citet{Sahlmann2011}, but a low significance fit to \textit{Hipparcos} data meant that the companion mass upper limit was enormous ($10.46~$M$_{\sun}$).

The orbital inclination is calculated in \textit{Gaia} DR3 as $i = 64.61 \pm 8.08^{\circ}$. This gives a resulting mass of $M = 86.12 \pm 5.76$~M$_{\textrm{jup}}$. This object has just been pushed over the HBMM threshold. The orbital period is only $15$~d (with orbital parameters $P$ and $e$ consistent between previous and current works), meaning the companion will receive considerable irradiation from the F-type primary star. Unless the companion is a very old BD that has slowly gained mass \citep{ForbesLoeb2019}, or is very rapidly rotating and has an increased HBMM \citep{Chowdhury2022}, it should be re-classified as a low-mass star.

The upper mass limit solely from \textit{Gaia} data is $1.42$~M$_{\sun}$ ($\sim1500$~M$_{\textrm{jup}}$), a tenth of that from \textit{Hipparcos}, exemplifying the improvement in the next generation of instruments. The lower limit in the \texttt{binary\_masses} table is quoted as $0.164$~M$_{\sun} ( \sim172$~M$_{\textrm{jup}}$) -- about twice the minimum mass derived from radial velocities and astrometry combined. Further study of this object is therefore required to find consistent mass measurements. Regardless, it is apparent from this work that the companion is actually a very low-mass star, capable of sustaining hydrogen burning, upon consideration of either the RV+NSS or \texttt{binary\_masses} results.

\subsubsection{HD\,105963\,A}\label{subsub:HD105963}
HD\,105963 has long been known as a wide-orbit binary with separation of 13.5~arcsec \citep[see][and references therein]{Grandjean2021}. During SOPHIE observations with a 3~arcsec fibre, \citet{Grandjean2021} detected a secondary component in the CCF, concluding that HD\,105963\,A is of type SB2. However, the authors could not fit an orbital solution to the RVs, so the secondary component of the inner binary remained un-characterised.

HD\,105963\,A (HIP\,59432) is detected by \textit{Gaia} to have significant deviation from a single-object astrometric model, with RUWE $\sim 34$. The NSS database  describes the orbital solution as having $P = 646.01 \pm 0.62$\,d, $e = 0.010 \pm 0.003$. This obviously cannot be attributed to the wide-orbit component, and provides a solution for the inner binary comprising HD\,105963\,A and a newly characterised companion HD\,105963\,C (naming in order of discovery, as wide-orbit companion has already been given designation B: \citealt{WashingtonDoubleStar,Gonzalez2023}). 

\begin{figure}
    \centering
    \includegraphics[width=0.90\columnwidth]{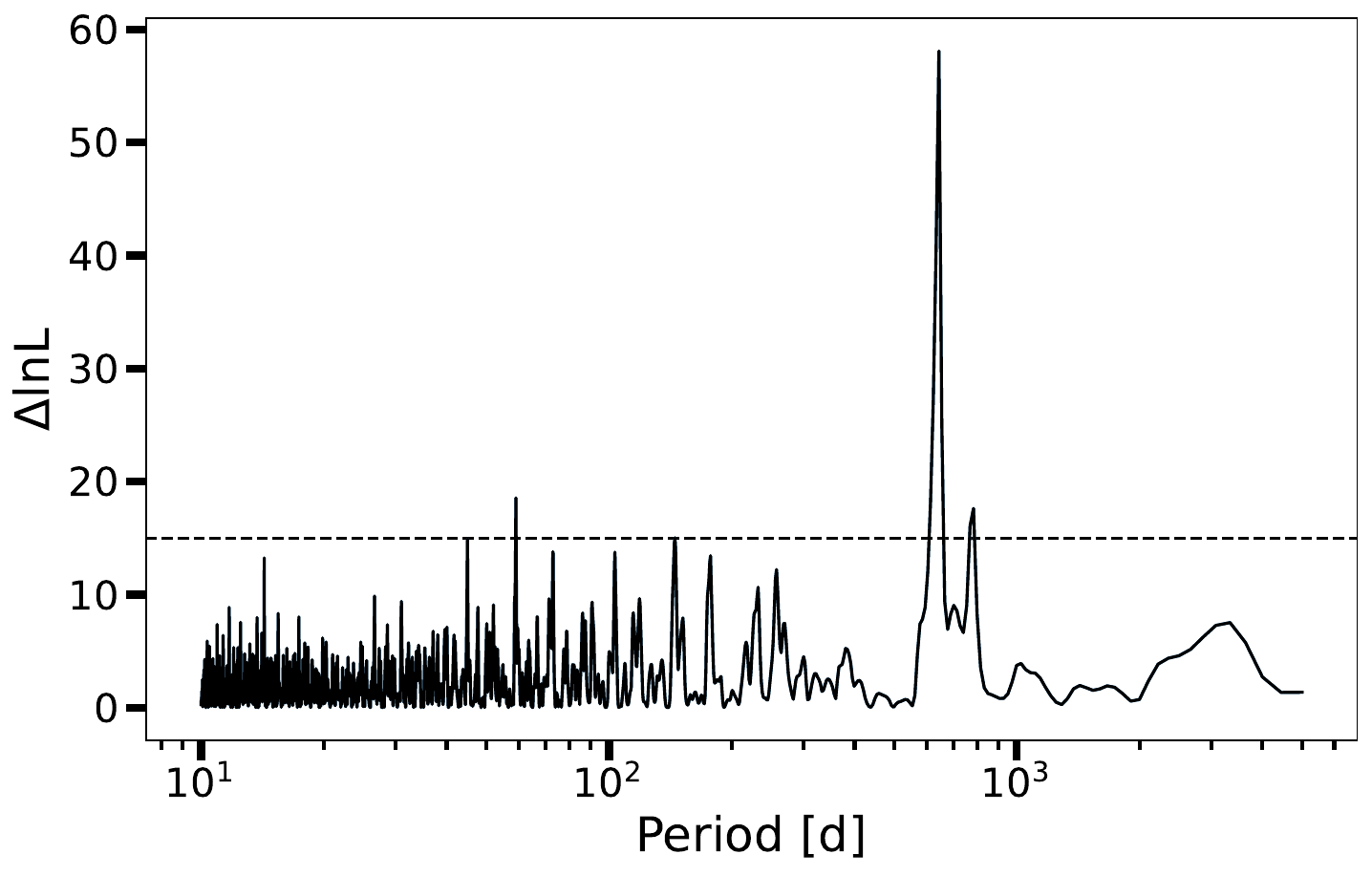}
    \hspace*{-0.60cm}\includegraphics[width=0.98\columnwidth]{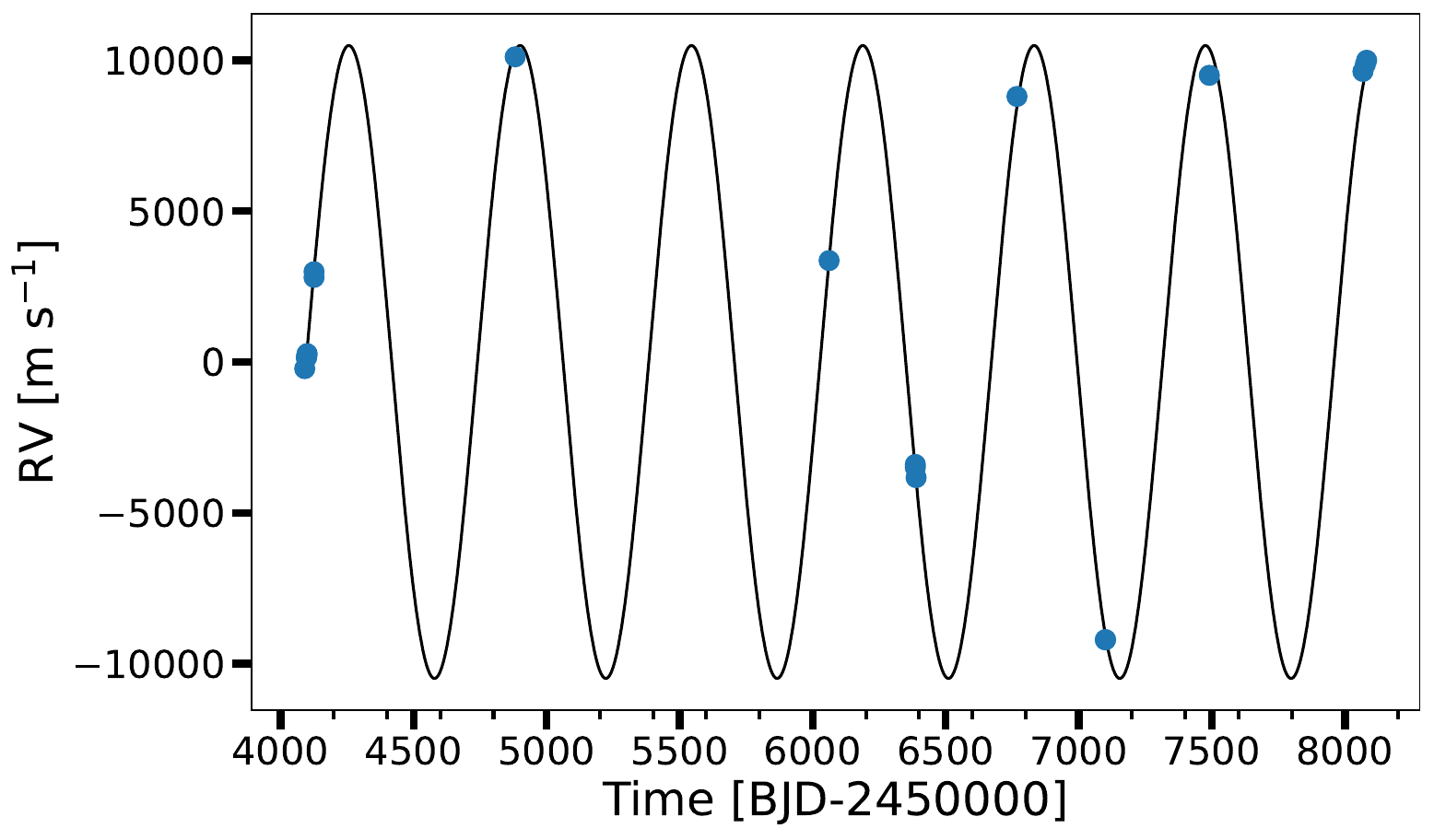}
    \caption{HD\,105963\,A. \textbf{Top}: MLP for archival SOPHIE data. The strongest peak is detected at $P = 643.90$~d. \textbf{Bottom}: A sinusoidal model fitted to the RVs, with parameters from the results of the MLP analysis \citep{ZK09}.}
    \label{fig:HD105963}
\end{figure}

With this solution from \textit{Gaia}, we can return to the RVs used by \citet{Grandjean2021}, and attempt to fit a Keplerian signal by using much more restrictive, informative priors on period and eccentricity based on the NSS information. The SOPHIE \citep{SOPHIE} RVs\footnote{SOPHIE archive RVs downloaded from \url{http://atlas.obs-hp.fr/sophie/} \citep{ELODIEArchive}.} were analysed with the \textsc{exo-striker} toolkit \citep{exostriker}. A maximum likelihood periodogram (MLP) search reveals a significant peak at $P=643.90$~d, and fitting a circular Keplerian we find a semi-amplitude of $\sim10$~km\,s$^{-1}$ (see Fig.~\ref{fig:HD105963}). This allows us to solve Kepler's third law and determine the RV predicted minimum mass $M\sin{i} = 567.7^{+24.1}_{-22.8}$~M$_{\textrm{jup}}$ (see Sect.~\ref{sec:method} for method). Using \textit{Gaia}-derived inclination ($i = 125.17 \pm 0.11^{\circ}$), the mass of this object is estimated to be $M = 664.4\pm 29.5$~M$_{\textrm{jup}}$.

The \texttt{binary\_masses} table also provides an estimation of mass for this companion. Through isochrone fitting for the primary component ($M_{\textrm{A}} = 0.927 \pm 0.051$~M$_{\sun}$), upper and lower bounds on companion mass are calculated to be [$0.42, 0.99]$~M$_{\sun}$ (alternatively, [$440, 1038$]~M$_{\textrm{jup}}$). Therefore, even with relatively weak constraints on the mass, this value is in agreement with that derived from the radial velocities, and we can confirm HD\,105963 is indeed a hierarchical triple star system.

\subsubsection{HD\,160508}
The brown dwarf candidate companion in orbit around HD\,160508 was discovered by \citet{Wilson2016}, with minimum mass $M\sin{i} = 48 \pm 3$~M$_{\textrm{jup}}$, and conservative maximum mass of 
$\sim1655$~M$_{\textrm{jup}}$ (from SOPHIE RV and \textit{Hipparcos} astrometry, respectively). The period and eccentricity are supported by results from \textit{Gaia} NSS, and are in agreement within the error bounds (Table~\ref{tab:NewTrueMasses}). The \textit{Gaia}-derived inclination is $i = 169.83 \pm 10.15^{\circ}$, therefore the mass of this object is estimated to be $M = 271.81\pm 268.95$~M$_{\textrm{jup}}$. As this error on orbital inclination includes configurations that are extremely close to face-on, the errors on the mass are also very large. The lower bound is below the minimum derived from RVs, so clearly needs to be treated differently. We can instead take the lower limit on mass as lower bound of $M\sin{i}$ ($45$~M$_{\textrm{jup}}$), solved with the least face-on inclination ($i = 159.68^{\circ}$). This results in a final spread of $M = 271.81~[129.58,540.76]$~M$_{\textrm{jup}}$. We can therefore conclude that even with large uncertainty bounds, this object is not a BD, but a low-mass star.

\subsubsection{HD\,191760}
The BD candidate orbiting HD\,191760 was first reported by \citet{Jenkins2009}, who used high precision RV to determine the orbital parameters. The orbital elements were reported as $P = 505.65 \pm 0.42$, $e = 0.63 \pm 0.01$, $K = 1047.8 \pm 38.7$. With host mass of $1.28$~M$_{\sun}$, the minimum mass of the companion was found to be $M\sin{i} = 38.17 \pm 1.02$~M$_{\textrm{jup}}$ -- placing it squarely in the brown dwarf desert.

\citet{Sahlmann2011} attempted to characterise the orbit of this object with \textit{Hipparcos} astrometry. They found inclination of $i = 15.4^{\circ}\,^{+21.0}_{-6.8}$, and mass of $M = 185.3^{+94.1}_{-93.4}~$~M$_{\textrm{jup}}$, before ultimately concluding that the orbit was of too low significance to effectively constrain the parameters. The astrometry did not reveal the signature of the companion reported by \citet{Jenkins2009}, and they considered the solution parameters invalid.

We have searched the \textit{Gaia} databases for information on HD\,191760. The NSS pipeline finds $P = 506.38 \pm 3.15$, $e = 0.625 \pm 0.044$, in excellent agreement with the solution from \citet{Jenkins2009}. The DR3 data has clearly detected the same companion, and we can analyse the orbit with \textsc{nsstools} to determine inclination and constrain the mass.

The updated inclination is now found to be $i = 158.96 \pm 5.99~^{\circ}$, and the mass can be solved as $M = 106.34 \pm 29.03$~M$_{\textrm{jup}}$. The lower bound of this mass lies around the HBMM limit ($\sim 77$~M$_{\textrm{jup}}$). Despite this, the companion is more than likely a VLMS, and should be capable of sustained hydrogen fusion.

\subsubsection{TYC~0173-02410-1}\label{subsub:TYC-0173}
A BD candidate from the SDSS-III Multi-object APO Radial Velocity Exoplanet Large-area Survey (MARVELS) survey, the companion to TYC~0173-02410-1 is one of ten discovered by \citet{Grieves2017}. Designated MARVELS-9\,b, orbital parameters were listed as $P = 216.5 \pm 1.8$~d, $e = 0.37 \pm 0.05$, with $M\sin{i} = 76.0 \pm 5.1$~M$_{\textrm{jup}}$ ($M_{\textrm{star}} = 1.06 \pm 0.10$~M$_{\sun}$).

Analysis of \textit{Gaia} DR3 data reveals a similar, but distinctly different orbital solution for the companion. The NSS table lists the orbital parameters as $P = 249.71 \pm 0.33$\,d, $e = 0.337 \pm 0.019$. 

\begin{figure}
    \centering
    \includegraphics[width=0.90\columnwidth]{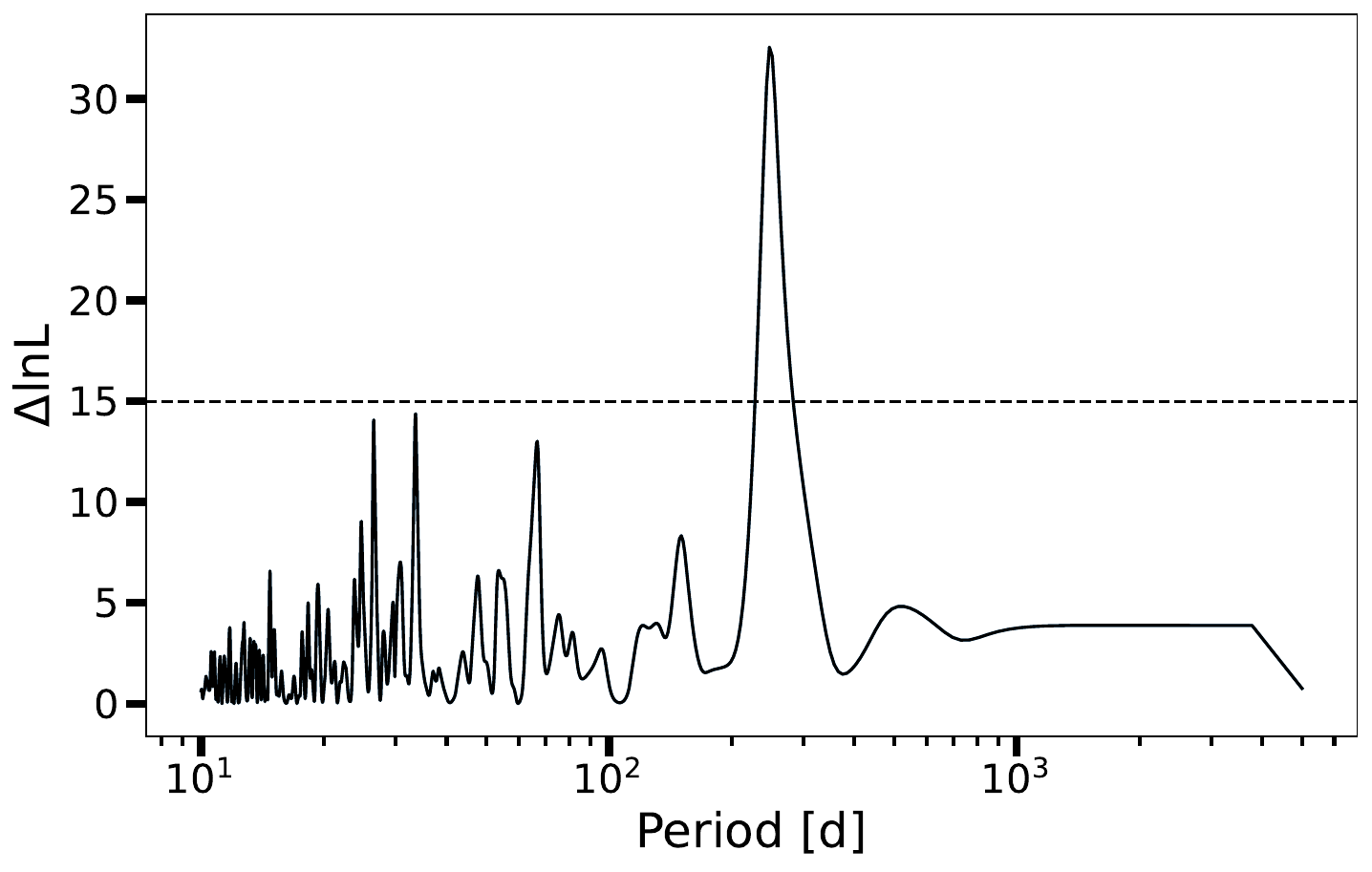}
    \hspace*{-0.60cm}\includegraphics[width=0.98\columnwidth]{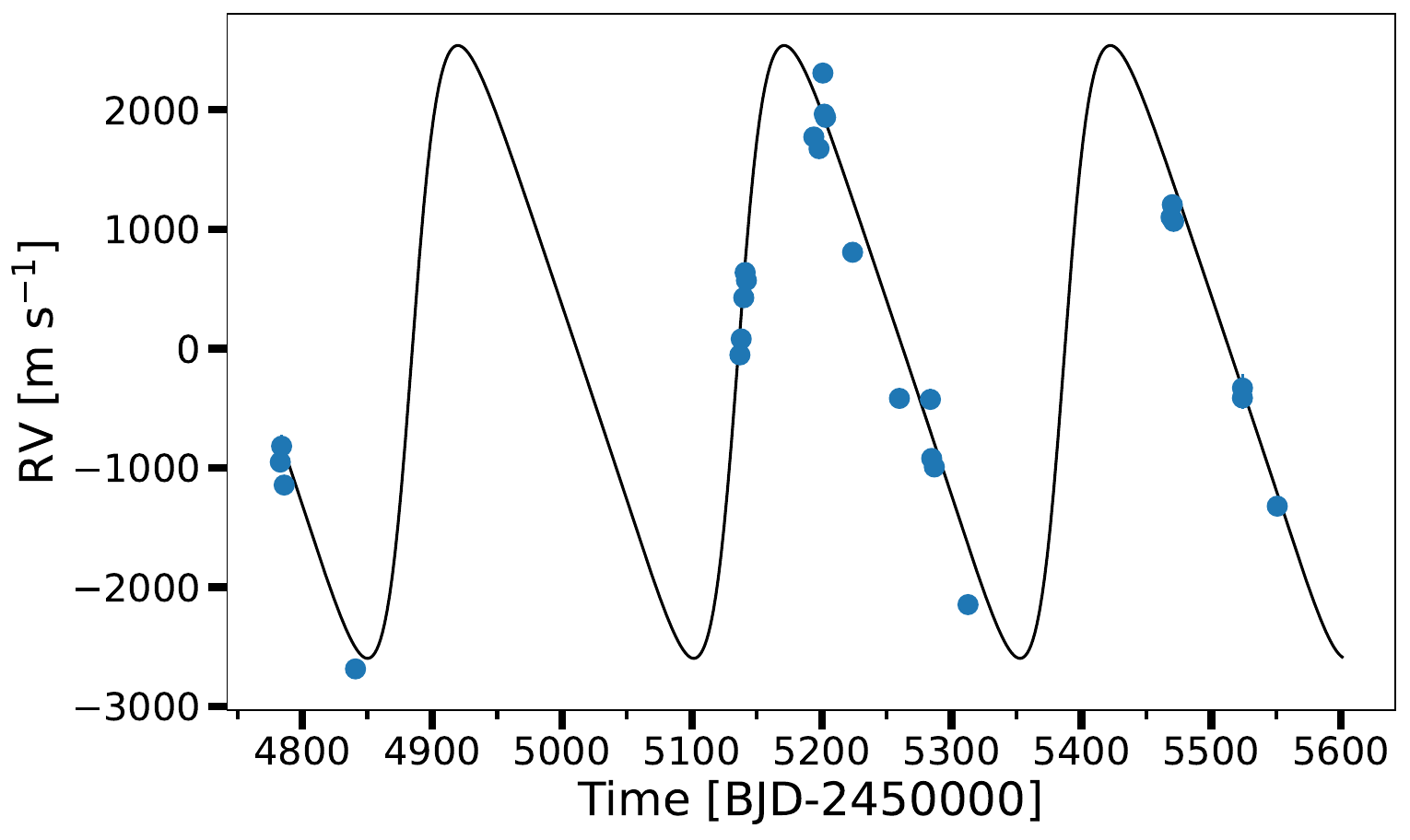}
    \caption{TYC~0173-02410-1. \textbf{Top}: MLP analysis. The strongest peak is detected at $P = 248.7$~d. \textbf{Bottom}: Keplerian orbital model fitted to the RVs.}
    \label{fig:TYC-0173-02410-1}
\end{figure}

We accessed the MARVELS RVs\footnote{Radial velocities for TYC~0173-02410-1 accessed within SDSS data directory at \url{https://data.sdss.org/sas/dr17/marvels/spectro/redux/v003.06/ASCII/GL273/}.} for this target, and performed an independent analysis. MLP routines find a signal period of $248.7 \pm 14.0$~d (Fig.~\ref{fig:TYC-0173-02410-1}), matching the NSS results well. Our results are different from that of \citet{Grieves2017} as we did not fit a linear trend before performing the period search. There is no strong justification for adding this linear trend, other than visual inspection of the RVs -- where a trend may be apparent, but is likely heavily impacted by the limited number of data points and sparse sampling. Additional confirmation that no linear trend is required is provided by the excellent match between our new solution and that from the NSS.  

Fitting to this updated period, we can determine $K$, and via numerical methods solving Kepler's third law, find an updated minimum mass of $M\sin{i} = 81.75 \pm 7.75$~M$_{\textrm{jup}}$ \citep[where we have used primary mass as reported in][]{Grieves2017}. Whilst being broadly consistent with the previous estimate, $M\sin{i}$ has been increased due to the longer period.

The derived inclination is found to be $i = 13 \pm 6^{\circ}$, meaning this orbit is almost face-on. The mass is therefore estimated to be $M = 362.86 \pm 169.25$~M$_{\textrm{jup}}$, with large errors propagated through from sizable inclination uncertainties, which have a large effect since the orbit is close to face-on. The object, previously straddling the hydrogen burning threshold, is now firmly placed within the stellar mass regime. This is broadly consistent with the wide mass range from the \texttt{binary\_masses} table, where the limits on the companion mass are [$265, 1097$]~M$_{\textrm{jup}}$. 

\subsubsection{GSC~03467-00030}
Similarly, GSC~03467-00030 was designated MARVELS-13 \citep{Grieves2017} after RV analysis found a companion on an orbit around the host with $P = 147.6 \pm 0.3$~d, $e = 0.50 \pm 0.04$, corresponding to a BD candidate with $M\sin{i} = 41.8 \pm 2.9$~M$_{\textrm{jup}}$ ($M_{\textrm{star}} = 0.96 \pm 0.09$~M$_{\sun}$)

\begin{figure}
    \centering
    \includegraphics[width=0.90\columnwidth]{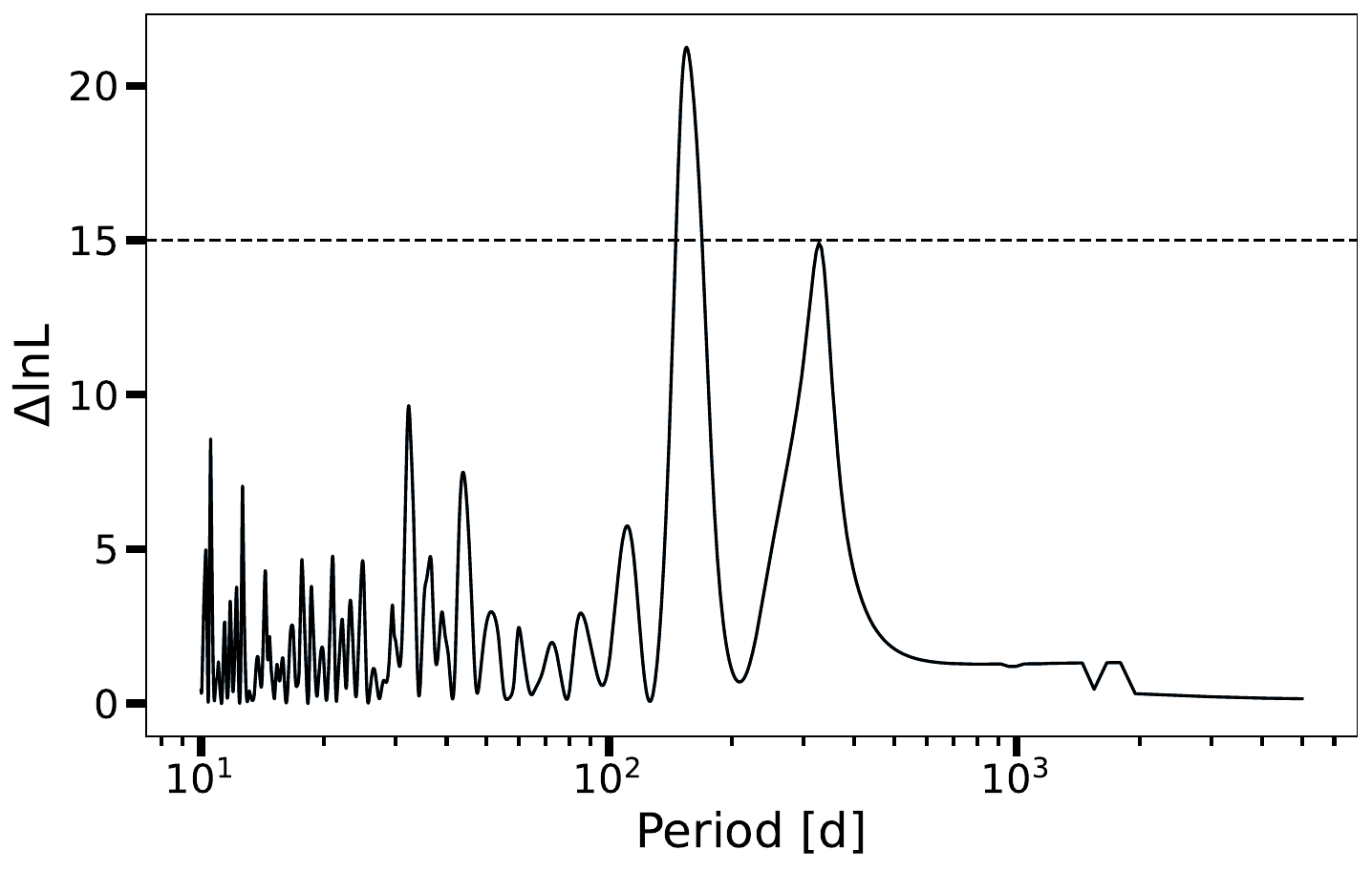}
    \hspace*{-0.60cm}\includegraphics[width=0.98\columnwidth]{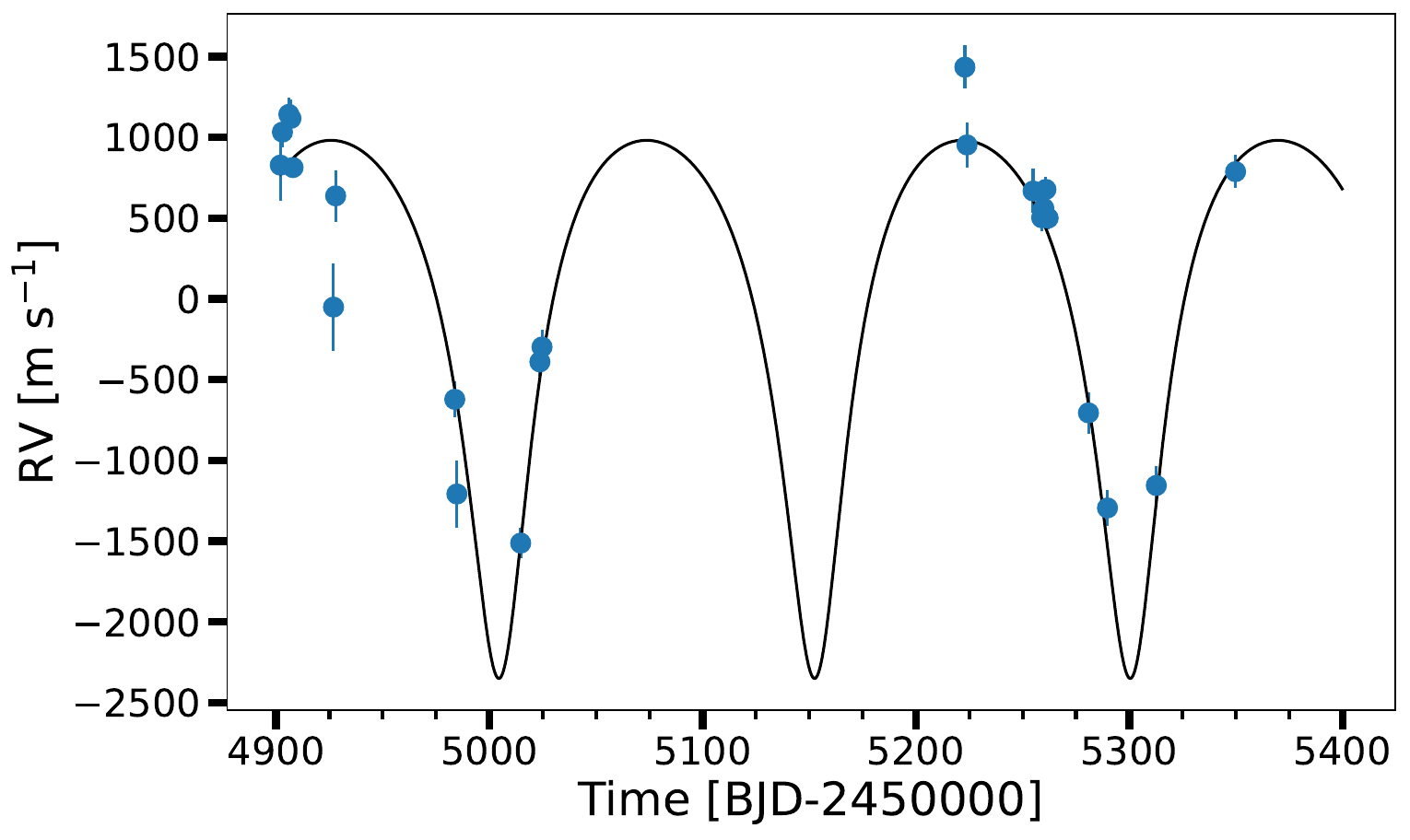}
    \hspace*{-0.60cm}\includegraphics[width=0.98\columnwidth]{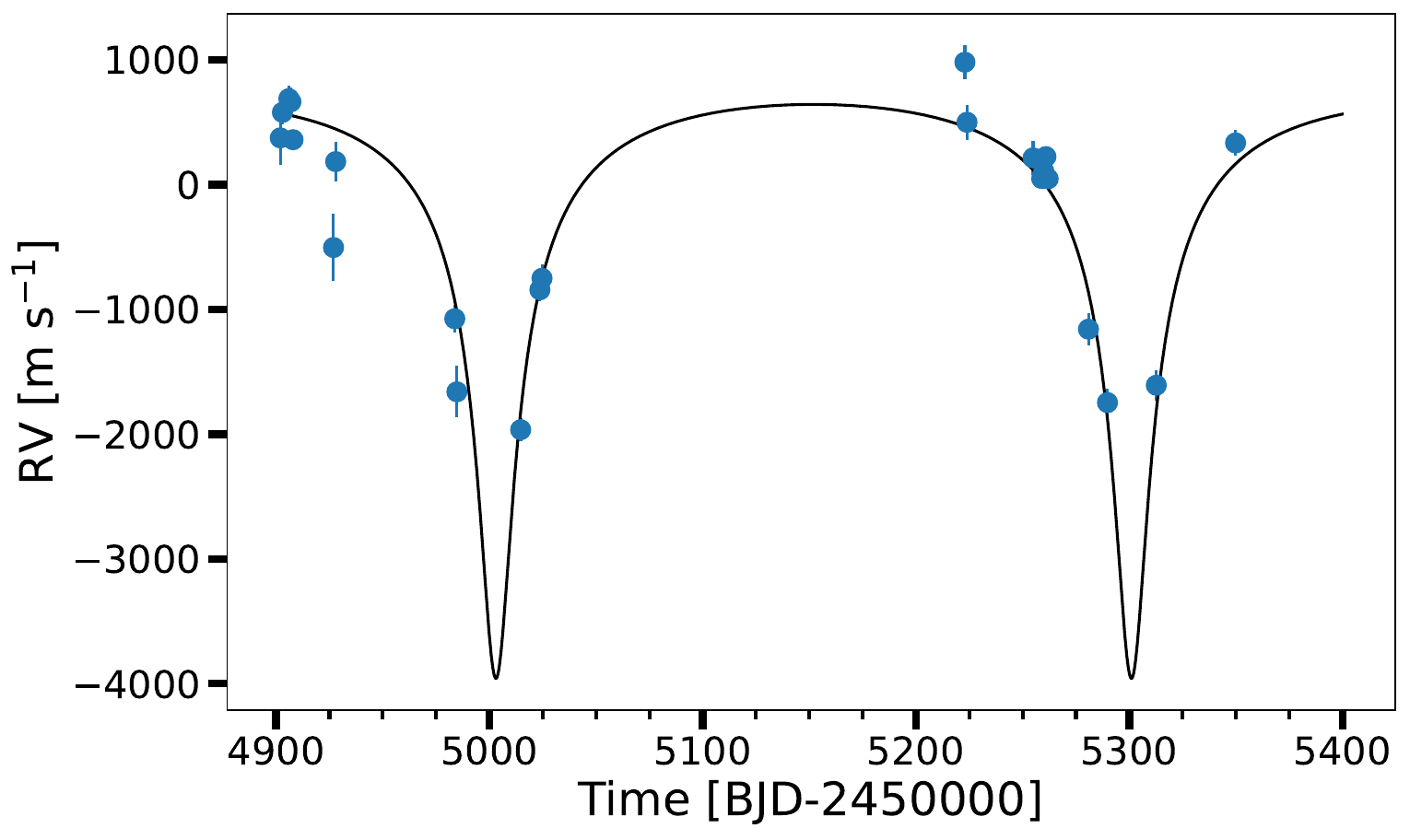}
    \caption{GSC~03467-00030. \textbf{Top}: Maximum likelihood periodogram analysis. The strongest peak is detected at $P = 155$~d, with the harmonic of this peak also detected with significant power ($\lnL \sim 15$). \textbf{Middle}: Full Keplerian model for the literature period fitted to the RVs. \textbf{Bottom}: Full Keplerian model for the \textit{Gaia} period fitted to the RVs.}
    \label{fig:GSC-03467-00030}
\end{figure}

\textit{Gaia} astrometry NSS reduction finds orbital motion at $P = 295.77$~d, twice the period of \citet{Grieves2017}. From the MARVELS RVs\footnote{Radial velocities for GSC~03467-00030 accessed within path \url{https://data.sdss.org/sas/dr17/marvels/spectro/redux/v003.06/ASCII/HAT-P-3/}.}, the MLP (Fig.~\ref{fig:GSC-03467-00030} top panel) shows power at both periods.  We find no significant residual power in the periodogram after fitting either signal, indicating both peaks in the periodogram arise from a single RV signal.

In some radial velocity studies, what appears at first to be an eccentric Keplerian solution can instead be made up of the superposition of two RV signals in 2:1 resonance \citep*{Anglada2010}. We attempted to fit two companion orbits to model the modulation seen in the RVs, at the two periods identified in Fig.~\ref{fig:GSC-03467-00030}. This proved unsuccessful, with no resonant solution combining these two periods fitting well to the data -- an eccentric solution at either period is vastly preferred.

On inspection of the statistical properties of the RV fits at each of the individual candidate periods (Fig.~\ref{fig:GSC-03467-00030} middle and bottom for the $\sim148$ and $\sim296$~d periods, respectively), the shorter period solution is preferred, but only by $\Delta \lnL = 7.258$. According to \citet{Kass+Raftery1995}, $\Delta \lnL \sim 7$ is the threshold for strong evidence, so one RV solution here is only \textit{just} preferable to the other. This is greatly affected by the sparse sampling, as additional RVs would clearly be beneficial to discriminate between solutions. On the other hand, \textit{Gaia} astrometric sampling will be far more uniform, and provide a longer observational baseline, with $\sim 2.8$~yr of data amassed by the time DR3 data was released. For this source there are 430 `good' astrometric observations (not down-weighted by measurement error), compared to 23 radial velocities spread over $\sim 1.2$~yr. Given this information, we choose to adopt the longer-period \textit{Gaia} derived solution, and recalculate RV parameters such as semi-amplitude and minimum mass.

We find the minimum mass is $M\sin{i} = 53.59^{+7.33}_{-5.50}$~M$_{\textrm{jup}}$, corresponding to an increase of $\sim 11.7$~M$_{\textrm{jup}}$. Inclusion of the inclination $i = 159.17 \pm 3.38^{\circ}$ allows mass to be estimated as $M = 150.68 \pm 31.16$~M$_{\textrm{jup}}$. However, this mass estimate is not close to that given in the \texttt{binary\_masses} table, where the mass bounds of the companion are [$494, 1007$]~M$_{\textrm{jup}}$. The cause of the inconsistency is unclear, and hints at the RV solution still being incorrect. Whilst further RV monitoring with improved cadence is required to truly determine what is orbiting with GSC~03467-00030, all mass estimates point to this object being stellar in nature, with masses well over the HBMM boundary.

\section{BD candidates from DR3}\label{sec:BD_bin_mass}
As mentioned previously, the NSS parameters are accompanied by mass estimates for each stellar pair with an astrometric solution in the \texttt{binary\_masses} table. For discussion on how these binary masses are calculated and potential sources of error, we direct the reader to \citet{GaiaStellarMulti2023}.
This mass information could therefore be used to expand upon the known sample of BD candidates, including those with no current RVs from ground-based surveys.

\citet{GaiaStellarMulti2023} describe the creation of a sub-sample by taking a cut on the companion masses, and find that 1843 sources have a companion with lower mass bound above a planet/BD transition defined at 20~M$_{\textrm{jup}}$ in their work. They identify a far smaller sample that also have upper limit below 80~M$_{\textrm{jup}}$, where orbital parameters are computed from both astrometry and spectroscopy. They however only tabulate the objects that match previously known companions from literature.

From our analysis in Section~\ref{sec:BDverification}, it is apparent that the \texttt{binary\_masses} upper \& lower limits can cover a very broad range. Therefore to assess any potential BD candidates, we chose to investigate companions with their entire uncertainty range lying inside the BD regime, i.e. those with with lower mass limit $>13$~M$_{\textrm{jup}}$, and upper mass limit $< 80$~M$_{\textrm{jup}}$. This rejects all objects that could feasibly be low-mass stellar companions.

Following these criteria we identify 24 additional companions from \texttt{binary\_masses}, with masses wholly in the BD range. Three of these are currently known: HD\,30246; HD\,92320; and HD\,132032. The first two are mentioned in \citet{GaiaStellarMulti2023}, and the third is already discussed in this paper (Section~\ref{subsub:HD132032}). Two more candidates are dropped for the following reasons: 
\begin{itemize}
    \item[--] HD\,340935: observations are listed as only single--lined spectroscopy, so will lack the required information to determine the mass, but instead calculate $M\sin i$; 
    \item[--] Gaia DR3 382531502736941440: this source has a clearly non-negligible flux ratio (close to unity), which is in conflict with with the predicted mass ratio. This object cannot be resolved by SIMBAD, and also possesses both a short period spectroscopic solution ($0.25$~d) and a long period astrometric solution ($1270$~d) in the NSS reduction. This source requires further study, beyond the scope of the present paper.
\end{itemize}

Consequently, our criteria selects 19 new candidate BD companions. These all have sufficient astrometric measurements to compute the inclination, and masses that lie completely within the BD regime (by design). These objects are tabulated in Table~\ref{tab:BD_bin_mass}. Interestingly, all candidate periods fall within the BD desert, and can preliminarily be considered in demographic studies -- vital when the BD desert appears to be becoming drier as a result of work in Section~\ref{sec:BDverification}. Further study into these systems will be required to assess the results from NSS processing in the BD regime.

\renewcommand{\arraystretch}{1.5}
\begin{table*}
	\centering
	\caption{Tabulated orbital parameters and masses for the new candidate companions discussed in Section~\ref{sec:BD_bin_mass}, identified from the \texttt{binary\_masses} DR3 database \citep{GaiaStellarMulti2023}.}
	\label{tab:BD_bin_mass}
	\hspace*{-0.9cm}\begin{tabular}{lccccccc} 
		\hline
		Host Star& $V_{\textrm{mag}}$ & $G_{\textrm{mag}}$ & $P_{\textrm{Gaia}}$~(d)  & $e_{\textrm{Gaia}}$ & $i~(^\circ)$ & $M_{\textrm{Host}}$~(M$_{\sun}$) & $M_{\textrm{BD}}$~(M$_{\textrm{jup}}$) \\
		\hline
        2MASS J04422788+0043376 & / & $17.27$ & $414.60 \pm 2.59$ & $0.202 \pm 0.042$  & $126.87 \pm 3.52$ & $0.1246^{+0.0484}_{-0.0478}$ & $51.614^{+12.202}_{-11.730}$ \\
        G 165-52 & / & $11.50$ & $156.22 \pm 0.20$ & $0.362 \pm 0.047$  & $136.14 \pm 2.84$ & $0.5549^{+0.0515}_{-0.0517}$ & $54.364^{+11.219}_{-11.391}$ \\
        GSC 04516-00523 & / & $11.86$ & $87.70 \pm 0.24$ & $0.388 \pm 0.073$  & $123.48 \pm 3.58$ & $0.6899^{+0.0560}_{-0.0488}$ & $68.205^{+11.750}_{-11.460}$ \\
        HD\,104289 & $8.07$ & $7.94$ & $1233.33 \pm 163.61$ & $0.377 \pm 0.0672$  & $117.28 \pm 3.44$ & $1.1534^{+0.0584}_{-0.0549}$ & $49.483^{+6.520}_{-6.615}$ \\
        HD\,115517 & $8.59$ & $8.46$ & $439.49\pm 4.94$ & $0.403 \pm 0.061$  & $47.62 \pm 5.09$ & $0.9692^{+0.0570}_{-0.0570}$ & $64.475^{+12.737}_{-12.563}$ \\
        HD\,156312\,B & $11.12$ & $10.67$ & $238.43 \pm 1.00$ & $0.239 \pm 0.064$  & $128.05 \pm 3.07$ & $0.9047^{+0.0570}_{-0.0603}$ & $66.519^{+10.539}_{-10.740}$ \\
        HIP\,117179 & $9.55$ & $9.37$ & $247.98 \pm 1.70$ & $0.416 \pm 0.077$  & $96.27 \pm 3.01$ & $0.9775^{+0.0596}_{-0.0620}$ & $44.197^{+4.786}_{-5.052}$ \\
        HIP\,60321 & $11.51$ & $10.86$ & $530.17 \pm 1.83$ & $0.340 \pm 0.014$  & $106.44 \pm 0.47$ & $0.6537^{+0.0553}_{-0.0497}$ & $68.260^{+10.256}_{-10.125}$ \\
        HIP\,75202 & $11.09$ & $10.97$ & $591.46\pm 5.16$ & $0.540\pm 0.039$  & $78.42 \pm 1.584$ & $0.8049^{+0.0549}_{-0.0527}$ & $69.009^{+9.994}_{-10.503}$ \\
        LP 498-48 & / & $15.89$ & $412.73 \pm 1.23$ & $0.279 \pm 0.029$  & $133.84 \pm 2.33$ & $0.1312^{+0.0484}_{-0.0484}$ & $46.424^{+10.086}_{-10.152}$ \\
        LSPM J1657+2448 & / & $16.18$ & $1182.34 \pm 58.66$ & $0.533 \pm 0.043$  & $109.87 \pm 1.10$ & $0.1420^{+0.0488}_{-0.0493}$ & [$36.464,~55.926$] \\
        LSPM J1831+4213 & / & $17.58$ & $324.05 \pm 7.62$ & $0.354 \pm 0.132$  & $58.71 \pm 6.84$ & $0.1288^{+0.0480}_{-0.0486}$ & $32.090^{+8.302}_{-7.636}$ \\
        TYC 3056-264-1 & $11.35$ & $10.96$ & $564.96 \pm 2.84$ & $0.487 \pm 0.025$  & $107.04 \pm 0.88$ & $0.7477^{+0.0497}_{-0.0529}$ & $64.727^{+10.158}_{-10.473}$ \\
        TYC 3873-761-1 & $10.49$ & $10.30$ & $526.19 \pm 3.26$ & $0.546 \pm 0.042$  & $108.04 \pm 1.40$ & $0.8885^{+0.0602}_{-0.0560}$ & $63.302^{+8.897}_{-9.128}$ \\
        TYC 7572-327-1 & $10.16$ & $10.01$ & $406.50 \pm 4.23$ & $0.476 \pm 0.054$  & $85.65 \pm 2.53$ & $0.9385^{+0.0582}_{-0.0628}$ & $70.921^{+8.612}_{-8.963}$ \\
        TYC 8321-266-1 & $10.30$ & $9.87$ & $413.89 \pm 5.94$ & $0.505 \pm 0.077$  & $96.91 \pm 5.17$ & $0.9572^{+0.0605}_{-0.0593}$ & $46.226^{+8.842}_{-8.644}$ \\
        TYC 9255-929-1 & $11.06$ & $10.68$ & $298.47 \pm 0.84$ & $0.229 \pm 0.073$  & $123.51 \pm 1.80$ & $0.8151^{+0.0508}_{-0.0533}$ & $61.629^{+12.060}_{-12.027}$ \\
        UCAC2 9182345 & $11.94$ & $11.51$ & $130.25 \pm 0.39$ & $0.439\pm 0.060$  & $59.11 \pm 2.87$ & $0.6829^{+0.0519}_{-0.0532}$ & $64.304^{+10.104}_{-10.394}$ \\
        UCAC4 302-050985 & $12.39$ & $11.73$ & $276.51 \pm 0.60$ & $0.374\pm 0.026$  & $110.64 \pm 1.57$ & $0.6649^{+0.0524}_{-0.0494}$ & $65.153^{+8.557}_{-8.712}$ \\
		\hline
	\end{tabular}
\end{table*}

\section{Implications for the BD desert}\label{sec:discussion}

\subsection{Period and Mass}
\begin{figure*}
    \includegraphics[width=0.80\linewidth]{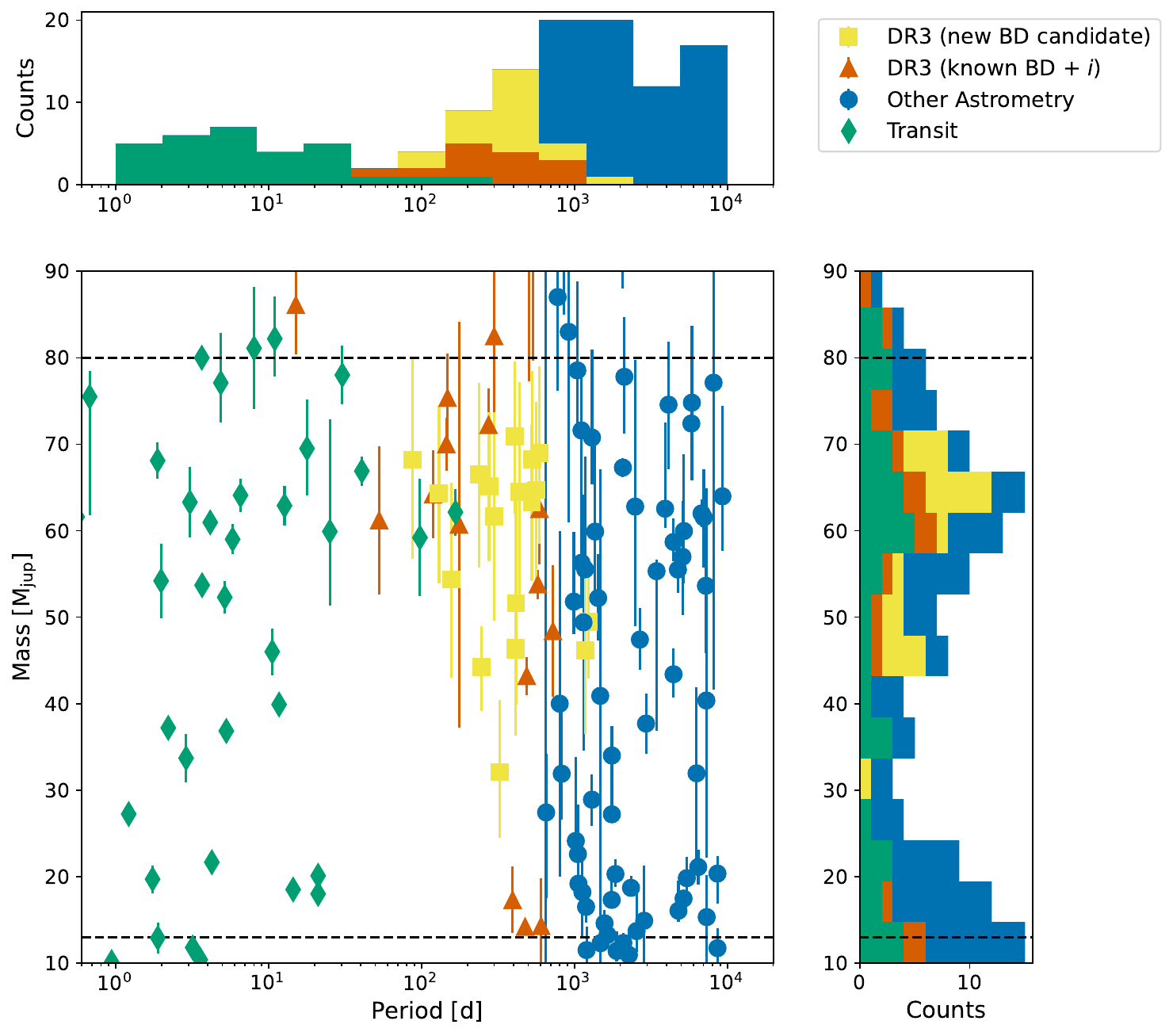}
    \caption{Masses of brown dwarf companions, plotted against orbital period. Histograms show the distributions over individual axes of the plot. Different methods of constraining mass are colour-coded and use differing shapes: transit observations are green diamonds; masses derived using other \textit{Hipparcos/Gaia} astrometry are blue circles; RV+NSS masses are orange triangles; and masses from \texttt{binary\_masses} are yellow squares.}
    \label{fig:P_truemass_methods}
\end{figure*}

\begin{figure}
    \centering
    \includegraphics[width=0.95\columnwidth]{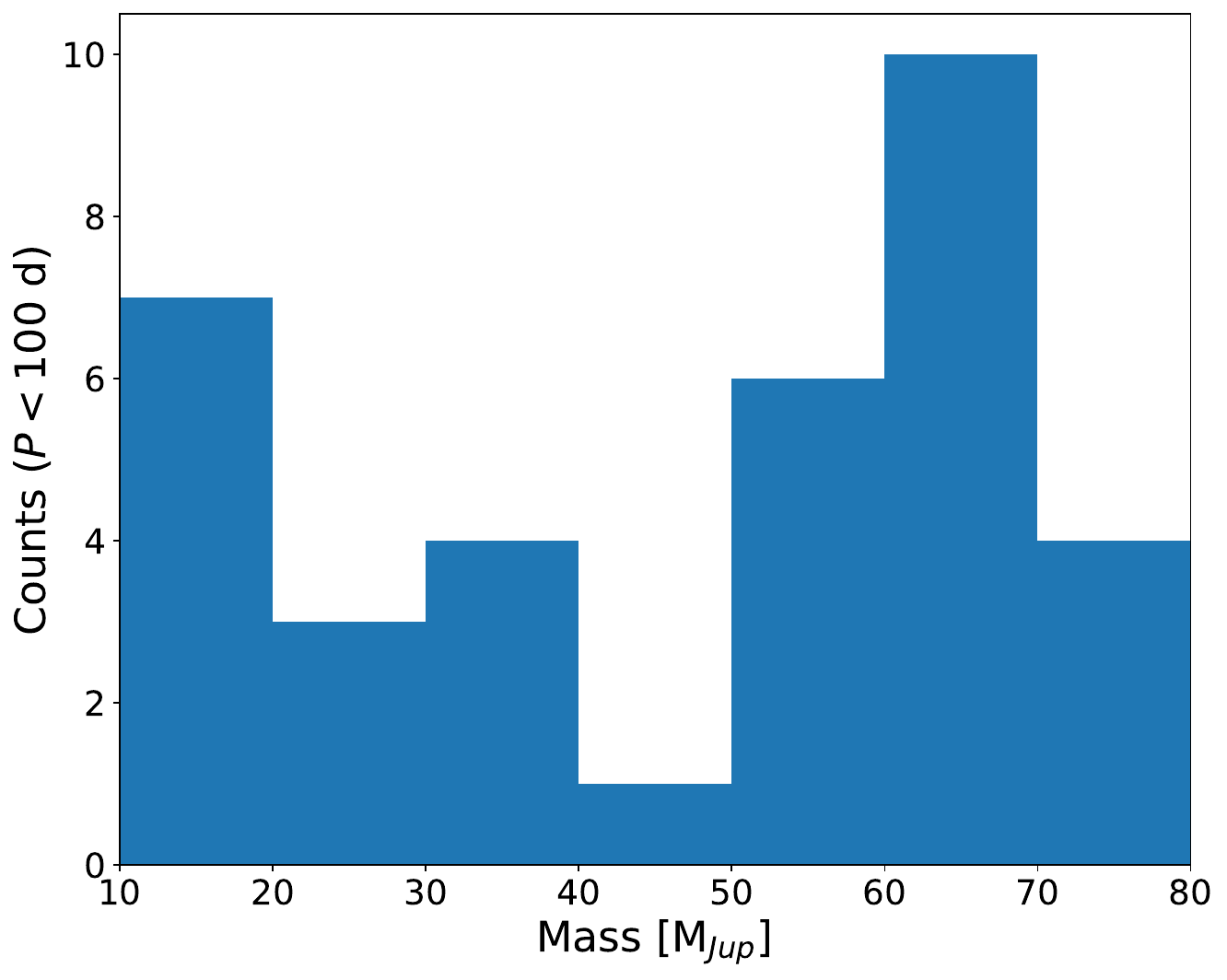}
    \caption{A histogram of BDs with confirmed masses in the BD region, for periods less than 100\,d. This helps us examine the "driest zone" \citep{MaGe2014}, indicating that there could still be a further deficit from 40--50~M$_{\textrm{jup}}$. This may represent the minimum of two different mass distributions, for either planets or stars \citep{GretherLineweaver2006}, but the low number of detections makes this difficult to quantify with available data.}
    \label{fig:Mhist_dryzone}
\end{figure}

Fig.~\ref{fig:MvsP} shows the period-mass space for all brown dwarfs in the desert. By removing those with only minimum mass information, an alternative plot is displayed in Fig.~\ref{fig:P_truemass_methods}. 
The region below 100~d is still more sparsely populated than longer periods (see period histogram in Fig.~\ref{fig:P_truemass_methods}), although recent BD transit detections have increased the number of known objects \citep{Carmichael2020,Grieves2021}. It is no longer as clear if the \citet{MaGe2014} driest region remains, with the inclusion of additional transiting BDs. There is still a region for $P<100$~d with mass between 40--50~M$_{\textrm{jup}}$ that only contains two objects, so the driest zone has perhaps become smaller (see Fig.~\ref{fig:Mhist_dryzone}, plotted for objects on periods less than 100\,d). There is however a clear lack of low-mass BDs between 100 and 1000\,d, in comparison to higher-mass BDs. This could be due to the detection methods used in creation of the sample. 

Selection effects in the sample will impact the overall demographic of the known BD desert. Beyond 1000\,d, there is no lack of low-mass objects: specifically, this region is populated by studies from \citet{Feng2022} and \citet{Xiao2023}, where $P>1000$~d was a requirement for their samples. Astrometry is most sensitive to long period orbits and massive objects, so will preferentially derive masses for these companions. Short period, low-mass objects are predominantly detected through transits, therefore the observed population is dependent on selection effects due to transit probability decreasing with increasing period. Additionally, transit observing campaigns can be further biased towards short periods as missions such as \textit{TESS} \citep{TessMission} only view each sector for 27 days on average, making periods longer than this difficult to detect.

DR3 results have begun to fill in parameter space for higher-mass BDs. The orange and yellow points in Fig.~\ref{fig:P_truemass_methods} show where \textit{Gaia} has enabled the mass to be calculated (through combining inclination with RV parameters or directly derived from the DR3 measurements, respectively).
There is a clear diagonal envelope bounding the \textit{Gaia} DR3 points. This is attributable to the fact that the astrometric sensitivity to a companion extends to lower masses as period increases. Due to the sensitivity limitations of \textit{Gaia} DR3 astrometry, we cannot comment on the lack of low-mass BDs with period on the order of $100$~d.

\begin{figure}
    \includegraphics[width=0.95\columnwidth]{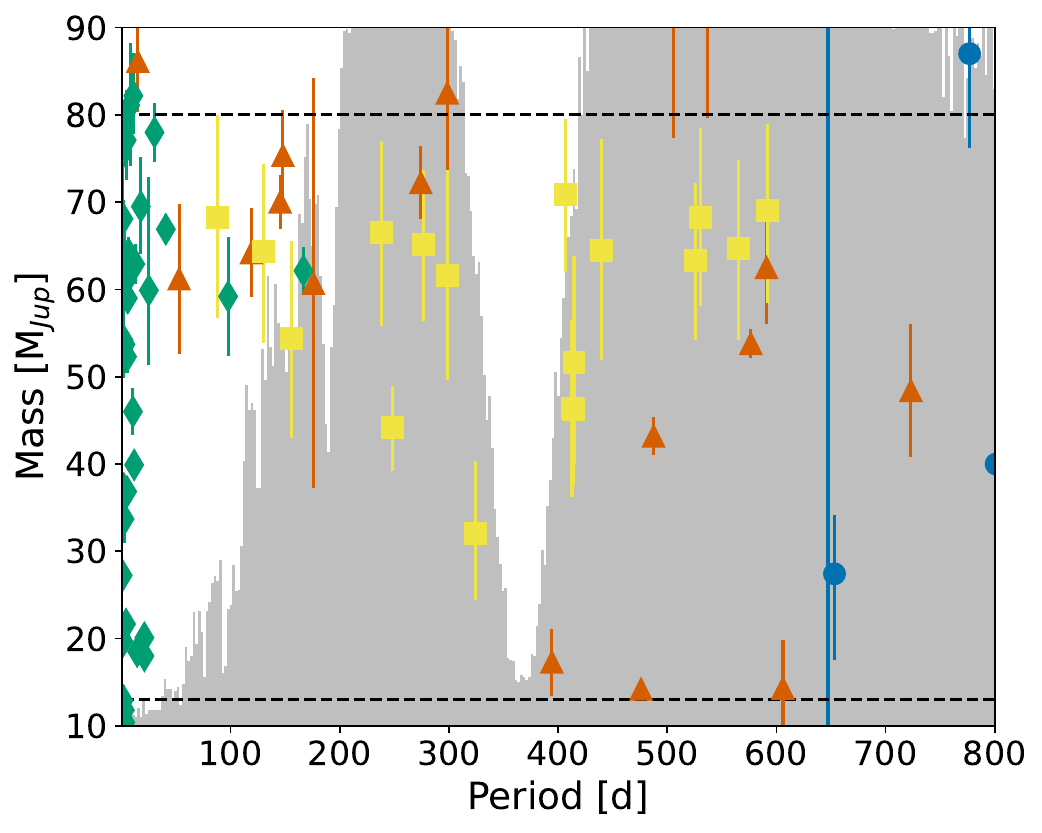}
    \caption{A section of the mass--period plane, showing the diagonal envelope bounding the \textit{Gaia} DR3 points (yellow and orange, with the same colours and shapes used as in Fig.~\ref{fig:P_truemass_methods}). Plotted in the background is an arbitrarily scaled histogram of DR3 astrometric solutions (inspired by \citealt{2023MNRAS.521.4323E}), identifying clear minima at $P<\sim50$~d and $P\sim365$~d. We see a gap in DR3 BDs lining up with the paucity of detections around 1~yr.}
    \label{fig:sensitivity}
\end{figure}

To assess how sensitive DR3 astrometry is to BDs at different periods, we have plotted a section of the mass--period plane in Fig.~\ref{fig:sensitivity}. This highlights the aforementioned diagonal envelope bounding the DR3 points, demonstrating the sensitivity changes at short periods. In the background of this Figure, we have plotted a histogram of all NSS astrometric detections. The histogram uses arbitrary scaling, as its purpose is to show periods that have a dearth of detections. Short periods are clearly highly suppressed, a consequence of the astrometry method and the small changes in position caused by a tight orbit. 
Interestingly, the histogram also shows a decreased number of detections around 1~yr \citep{2023MNRAS.521.4323E}. From our overlaid DR3 points, it is clear this is also observed in our dataset. The missing detections around 1~yr are caused by systematic effects in the way \textit{Gaia} measures positions. At L2, the spacecraft completes an orbit around the Sun in the same length of time as Earth, and any slight changes in stellar positions will be difficult to separate from annual motion around the Sun.

\citet{Kiefer2021} observed an empty region of the parameter space between masses of 20--85~M$_{\textrm{jup}}$, and $0<P<100$~d (their fig~16). We find that region to be sparsely populated, mostly only by BDs discovered through transit photometry. There is a significant lack of BDs in \textit{Kepler} results \citep{KeplerMission,Herald2022}, agreeing with the low number of short period, transiting BDs known in general. From the log-spaced period histogram in Fig.~\ref{fig:P_truemass_methods}, we see that the occurrence of BDs does indeed seem to transition beyond periods of $>100$~d, where the desert becomes `wetter'. 
However this impression could be entirely or partly caused by the selection effects, rather than revealing the underlying demographic distribution of BDs in binaries.
Many masses come from astrometry, where detections extend to lower masses at long periods, so the large number of low mass BDs with long periods could be due to sensitivity limitations of instruments, reductions, and software (see \citealt{Xiao2023} for discussion on the $>1000$~d period requirement of \textsc{orvara}).

\citet{Sahlmann2011} used \textit{Hipparcos} astrometry to break the $\sin{i}$ degeneracy in RV minimum masses, and found two populations of BDs, with low-mass from $\sim13-25$~M$_{\textrm{jup}}$, and high-mass BDs in the zone above $45$~M$_{\textrm{jup}}$. They concluded that the low-mass BDs were an extension of the planet distribution function, a continuation of a large-planet-mass tail. \citet{MaGe2014} suggested that BDs lighter than $42.5$~M$_{\textrm{jup}}$ form through planetary disc gravitational instability, whereas heavier companions would form like binaries through stellar cloud fragmentation. Recent studies corroborate this by finding a clear valley in the mass distribution near $40$~M$_{\textrm{jup}}$ \citep{Feng2022,Xiao2023}.

In the mass histogram of our Fig.~\ref{fig:P_truemass_methods}, there also appear to be two distinct populations, with a minimum between them at around 30--35 Jupiter masses. The two driving populations behind this are objects just over the planet--BD threshold (with many planet candidates `promoted' to BD status by \citealt{Xiao2023}), and DR3-based results clustering around 40--70~M$_{\textrm{jup}}$. Results here are similar to previous works, indicating that DR3 data aren't introducing the observed minimum of the mass distribution due to \textit{Gaia} selection effects. The empty region of the mass--period plane for low-mass BDs on periods of order 100~days could change the mass distribution if eventually filled in, but is unlikely to be drastically different to the currently observed arrangement.

The majority of RV-detected planets more massive than Jupiter were observed to be concentrated around $P\sim500$~d \citep{Liu2008}. 
We have searched for extra-solar giant planets (EGPs; $>1$~M$_{\textrm{jup}}$) in exoplanet databases. Restricting our search to those planets with only a minimum mass measurement, we also find a peak in the period distribution at $\sim500$~d. These EGPs could be awaiting astrometric treatment to change their classification from planets to BDs -- future Gaia data releases may be able to assist in populating this empty region of low-mass BDs with periods between 10s and 1000s of days. 

With available data we find that the dichotomy observed is likely to be a real phenomenon, as do studies in literature \citep{Sahlmann2011,Feng2022,Xiao2023}.

\subsection{Mass and Eccentricity}
The eccentricity and mass distribution of BD companions was used by \citet{MaGe2014} to determine a transition in the two populations, with a threshold at $42.5$~M$_{\textrm{jup}}$. Below this, BDs were observed to fit in with planetary eccentricity distributions, and over the transition, fit with stellar binary populations \citep{Halbwachs2003}. 

Low-mass BDs are hypothesised to follow a trend of decreasing eccentricity with increasing mass. BDs formed in the protoplanetary disc are likely to have been pumped to higher eccentricities by other companions, a feat that is easier to achieve the lower the mass of the object \citep[see the planet-planet scattering models of][and fig.~5 in \citealt{MaGe2014}]{FordRasio2008}.

\citet{Xiao2023} find a relatively empty eccentricity valley bounded by the upper profile of the eccentricity distribution (their fig.~14\,b). The maximum eccentricity in each mass range goes from $e\sim0.9$ at planetary masses, to $e=0.6$ at $M = 35$~M$_{\textrm{jup}}$, then back to $e\sim0.8$--$0.9$ for high-mass BDs $M > 42.5$~M$_{\textrm{jup}}$.  Their eccentricity valley is by the side of the transition mass found by \citet{MaGe2014}.

\begin{figure*}
    \centering
    \includegraphics[width=0.90\linewidth]{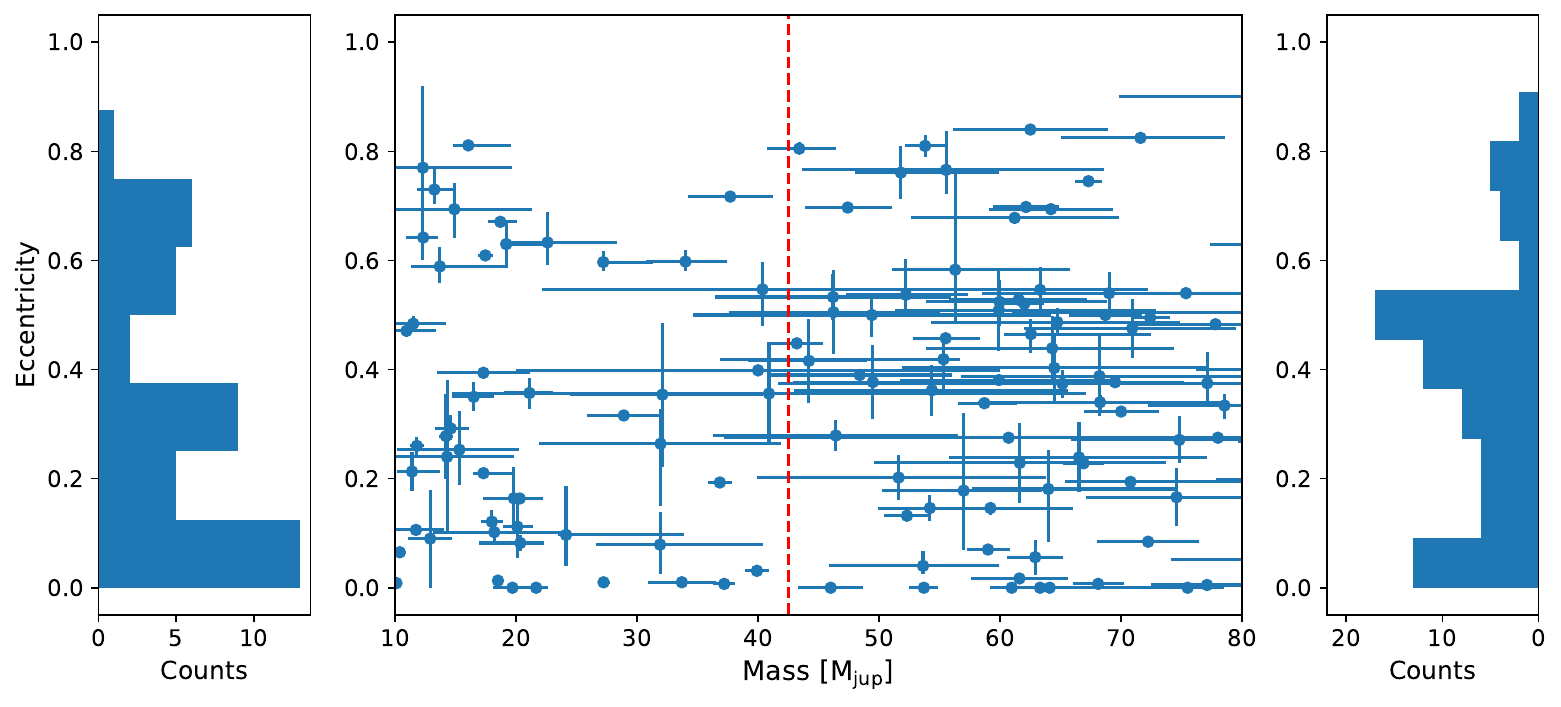}
    \caption{Companion orbital eccentricity plotted against mass. Histograms are plotted on either side for low-mass (left) and high-mass (right) BDs. The red dashed vertical line denotes the $42.5$~M$_{\textrm{Jup}}$ split.}
    \label{fig:MvsE}
\end{figure*}

To analyse the same phenomenon with our sample, we have plotted the mass-eccentricity plane in Fig.~\ref{fig:MvsE}. Our sample includes host stars (e.g. M-type stars from \textit{Gaia} results in Section~\ref{sec:BD_bin_mass}) that would be removed by the FGK requirement in works such as \citet{Kiefer2021} and \citet{Xiao2023}. These were left in to observe if changes to the populations would be immediately apparent with their inclusion.

We observe a similar mass-eccentricity distribution (Fig.~\ref{fig:MvsE}) to those in either \citet{MaGe2014} or \citet{Xiao2023}. Our distribution has either an eccentricity valley centered on roughly the same mass and eccentricity as seen in \citet{Xiao2023}, or a trend of decreasing maximum eccentricity with increasing mass for low-mass BDs -- from $e\sim0.8$ for the lowest mass BDs, down to $e\sim0.6$ at $M = 35$~M$_{\textrm{jup}}$. This downward trend would continue up to the low/high mass BD transition at $42.5$~M$_{\textrm{jup}}$, were it not for one outlier at $M \sim 37$~M$_{\textrm{jup}}$ and $e\sim 0.7$. Without this BD included, our result would look strikingly similar in form to the result in \citet{MaGe2014}. The outlier orbits HD\,122562, a G5 star, so has not arisen from our inclusion of some M-type stars. 

To determine whether the split at 42.5~M$_{\textrm{jup}}$ is statistically significant or not, we have used the Kolmogorov--Smirnov (K--S) test to compare companions in two groups: $13< {\frac{{M}}{{\textrm{M}_\textrm{jup}}}}  <42.5$ and $42.5<{\frac{{M}}{{\textrm{M}_\textrm{jup}}}}<80$. The two-sample K--S test calculates the probability, or $p$-value, that the two sets are drawn from the same distribution \citep{Hodges1958}. We find $p = 0.09$ for all the stars in our sample with both mass and eccentricity estimates. It is normal to reject the null hypothesis where the probability is less than 5 per cent ($p=0.05$).
Despite being above this 5~per~cent threshold, the low $p$-value is still indicative of inherent differences in the two populations with masses either greater than or lesser than $42.5$~M$_{\textrm{jup}}$. If we try different values for the mass split threshold value, we find a minimum $p$-value at $40$~M$_{\textrm{jup}}$ of $\sim5$~per~cent. It does indeed seem that there is some form of split around the mass threshold identified by \citet{MaGe2014}. This is likely to arise due to two different formation channels creating the observed companion orbital parameters.

We have also attempted to compare the eccentricity distributions with those of detected exoplanets (\href{http://exoplanet.eu}{\texttt{exoplanet.eu}}), or spectroscopic binaries \citep{SB9catalogue}.
Many companions are detected on low-eccentricity orbits, which could be impacted by observational biases. Short period orbits are preferentially detected in both transit and RV studies, and proximity to the host star increases the chance of a circularised companion orbit. In either of the aforementioned catalogues, or the BD desert sample, this will skew the observed detection frequencies. To assess similarities, we have applied the two sided K--S test, between planets and low-mass BDs, or high-mass BDs and stellar binaries \citep{Xiao2023}, but adding a cut on orbital period for $P>100$~d, in an attempt to avoid biases and selection effects.

\begin{figure}
    \centering
    \includegraphics[width=0.95\columnwidth]{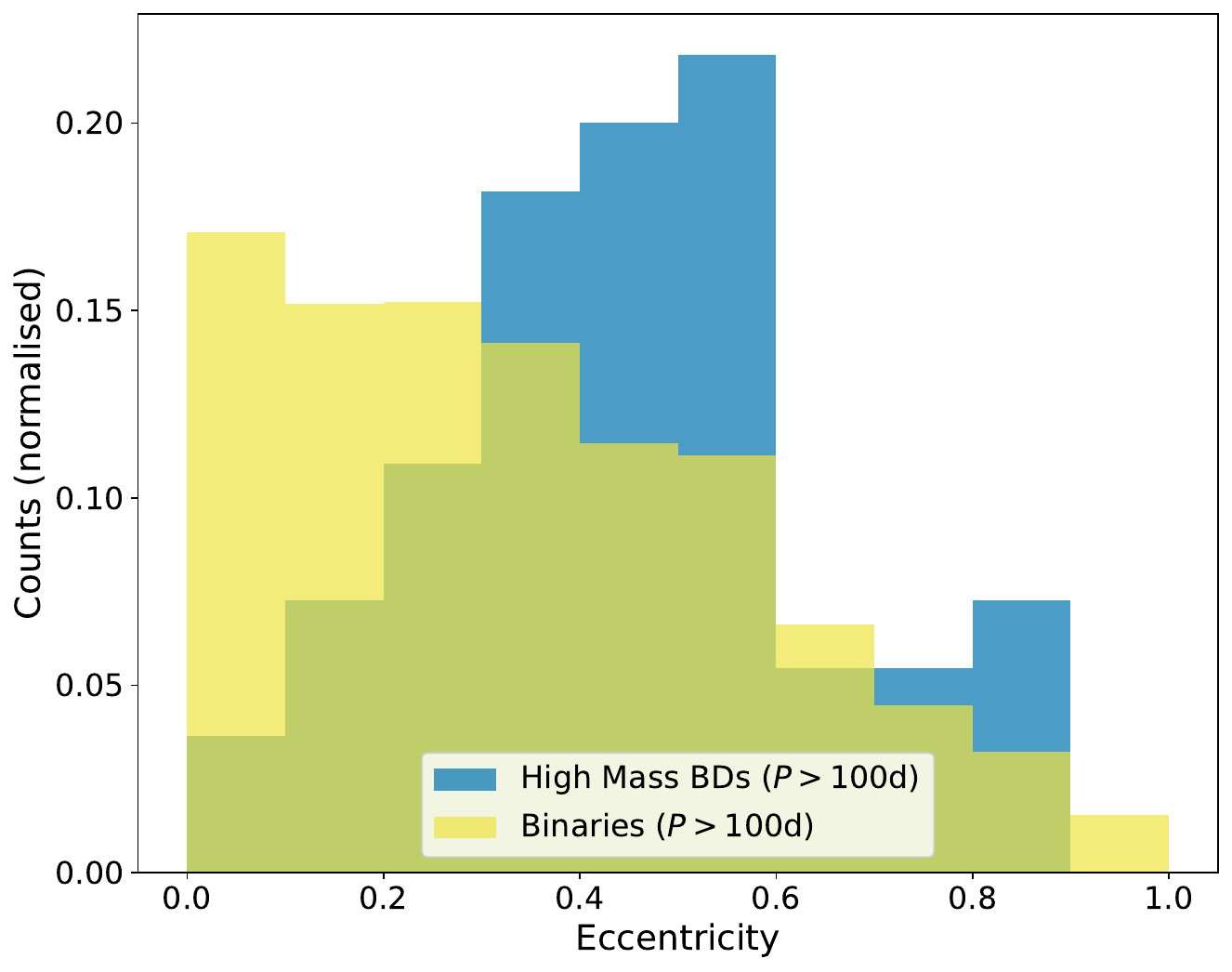}
    \caption{Histograms showing the differing normalised eccentricity distributions for high-mass BDs and spectroscopic binaries, with orbital periods greater than 100 days. We see an increased frequency of moderate-to-high eccentricities in BDs compared to the stellar binary sample.}
    \label{fig:highM-binaries-ecc}
\end{figure}

The eccentricities of low-mass objects could be drawn from the same distribution as planets, with $p$-value of $8$~per~cent indicating we have insufficient information to reject the null hypothesis of a single over-arching sample. This could be related to the low number of detected companion BDs, and hopefully can be re-addressed in the future. However, in disagreement with \citet{Xiao2023}, we find that the high-mass BDs are not drawn from the same sample as stellar binaries ($p = 0.0002$). Again, this could be due to the relative number of objects in the samples, but could also be caused by an over-abundance of high-mass BDs observed between eccentricities of $\sim0.35$ and $0.55$. This is highlighted by the histograms plotted in Fig.~\ref{fig:highM-binaries-ecc}.

\subsection{Period and Eccentricity}

The period-eccentricity plane of BDs was studied previously by \citet{MaGe2014}, but has not been included in more recent works assessing the BD desert. We can examine the plot again, but using only companions with a constrained mass, rather than only a minimum. 

\citet{MaGe2014} divided their sample into the two familiar groups, of high-mass and low-mass BDs split at $42.5$~M$_{\textrm{jup}}$. They found that high-mass BDs were consistent with the stellar binary circularisation limit of $\sim12$~d, and that there was a total depletion of high-mass BDs for $300<P<3000$ and $e<0.4$, not observed in low-mass BDs. Performing the 2D version of the K--S two-sample test, \citet{MaGe2014} found a probability of $1.7$~per~cent that low-mass and high-mass BDs had been drawn from the same parent distribution, providing more evidence that these two sub-sets formed through differing processes.

We have replicated the previous analysis, but this time including only companions with a constrained mass. We find a very similar 2D K--S\footnote{The two-dimensional version of this statistical test was performed using the \textsc{ndtest} code, freely accessible at \href{https://github.com/syrte/ndtest}{\texttt{github.com/syrte/ndtest}}.} $p$-value of $3.7$~per~cent, allowing us to also reject the null hypothesis that the two sub-sets are drawn from the same parent distribution. 

\begin{figure}
    \centering
    \includegraphics[width=0.99\columnwidth]{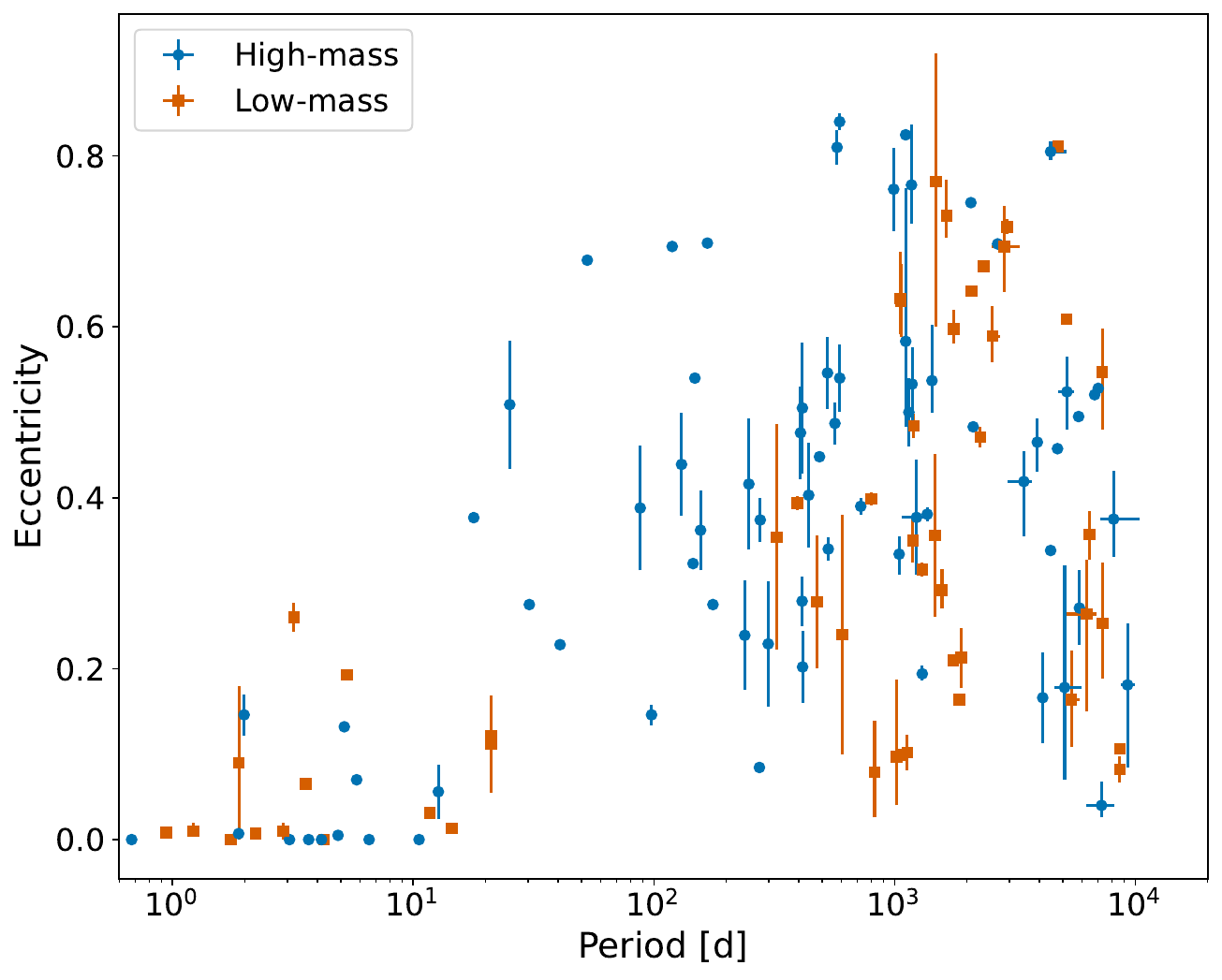}.
    \caption{Eccentricity plotted against period for stars with derived mass estimates. The two sets of high/low-mass have been split at $42.5$~M$_{\textrm{jup}}$, as in \citet{MaGe2014}. An empty region of circular orbits at periods of order 100~d is immediately apparent. Low-mass BDs are represented by orange squares, and high-mass BDs are represented by blue circles.}
    \label{fig:PvsE}
\end{figure}

However, the similarities between our study into period \& eccentricity and that of \citet{MaGe2014} end there. We find a non-negligible number of high-mass BDs on long period orbits with eccentricity below $0.4$ (see Fig.~\ref{fig:PvsE}). We also observe three high-mass BDs on orbits with $P<12$~d that are not completely circular, at odds with the circularisation limit seen in stellar binaries. These are:
\begin{enumerate}
    \item the companion to CoRoT-33 (G9V), a $59$~M$_{\textrm{jup}}$ transiting BD on a $5.8$~d orbit. The eccentricity is small but non-negligible, where $e=0.07 \pm 0.0016$ \citep{Csizmadia2015};
    \item The late F-type star EPIC 212036875 hosts a transiting BD of $52$~Jupiter masses on a $P=5.17$~d, $e=0.132$ orbit \citep{Carmichael2019};
    \item AD 3116/ EPIC 211946007 is comprised of a transiting BD orbiting a mid-M-dwarf. The BD has a $1.983$\,d period, and $e=0.146$ \citep{Gillen2017}. As this BD has such a low-mass host, it could remain in an eccentric orbit for longer than if it had a massive host. The system is found in a sub-Gyr age cluster, so is relatively young and may circularise over time.
\end{enumerate}

Our analysis has also identified a depleted region, for both high and low-mass BDs, of low-eccentricity/circular orbits centred on $P\sim100$~d. There are clearly orbits with low eccentricity observed at both short and long periods, bounding this `valley'. 

This distribution however may also suffer in the same way the period-mass plane does. 
Selection effects mean we cannot easily detect 
low-mass BDs at periods of order a few hundred days: astrometry is limited to longer periods for low-mass objects, and transits are mostly confined to shorter periods. 
Hence an undetected population of low-mass objects 
may be a missing piece of the puzzle, and fill in this region of the period-eccentricity plane.

\subsection{Mass and Metallicity}
In recent literature discussing the BD desert, many have linked metallicity to companion formation scenarios. There are multiple competing theories for the formation of large companions \citep{MaGe2014}. They are thought to either form in the protoplanetary disc around the host star, or through fragmentation of the pre-stellar molecular cloud into multiple objects -- parent star and companion, that then both contract to form individual bodies. The formation from a fragmented cloud of dust and gas is known as gravitational instability (GI), as perturbations cause over-dense regions that then contract under self-gravity.

Formation within the disc itself is accomplished through one of two mechanisms: either core accretion (CA), where colliding clumps of dust stick together to form planetesimals that then grow by consuming other material from the protoplanetary disc; or disc instability \citep{Boss1997} -- a form of GI where a massive disc fragments due to outside influence, or torques as the disc rotates, and the remaining clumps of material then contract under self gravity in a manner similar to that of molecular cloud GI. CA is likely dominant in metal-rich discs, where the metal core of a giant planet can efficiently coalesce \citep{Alibert2005}. Alternatively, disc instability would allow similar formation rates in both metal-rich and metal-poor protoplanetary environments \citep{Boss1997}.

Giant planet occurrence was found to be strongly correlated with host star metallicity \citep{Johnson2010}, leading to the conclusion that these objects predominantly form through the CA mechanism. In contrast, \citet{MaGe2014} found that BDs did not follow this correlation, supporting the hypothesis that low-mass BD companions in protoplanetary discs instead formed via the disc--instability mechanism \citep{Rice2003}.

\citet{MaldonadoVillaver2017} also showed that BD hosts do not follow the giant planet metallicity correlation. They concluded that CA may form low-mass BDs in metal-rich discs, and that low-mass BDs in metal-poor discs would be driven by disc instability - with the two mechanisms working in tandem to fill in the population. However, other works have studied mass thresholds for where each mechanism will be able to form a companion. \citet{Santos2017} found that objects $>4$~M$_{\textrm{jup}}$ would form through GI (either in the disc, or of a molecular cloud), rather than CA. \citet{Schlaufman2018} derived a limit of $10$~M$_{\textrm{jup}}$, above which companions would form via  some sort of gravitational instability instead of core accretion. Both of these works find CA is only possible below the commonly used DBMM, and this backs up the observations rejecting a metallicity-mass correlation for the BD regime -- as disc instability is theorised to form a companion regardless of the host star's metallicity.

In the recent study by \citet{Xiao2023}, they consider companions with masses ranging from those of planets up to those of low-mass stars to assess the BD desert. To investigate metallicity distributions, the sample is restricted to FGK host stars over 0.52 solar masses, to ensure accurate derivation of atmospheric parameters. In their fig.~14\,(a), \citet{Xiao2023} assert that they observe the BD desert to be located in a transition region between giant planets and low-mass stellar binaries. Planets preferentially orbit hosts with super-solar metallicity ($0.09 \pm 0.19$ dex), whereas stellar binaries are found to have sub-solar metallicity spanning a larger range ($-0.07 \pm 0.27$ dex). The sample is split into four groups: $M <13$; $13<M<42.5$; $42.5<M<80$; $M>80$~(all in units of M$_{\textrm{jup}}$). They find significant evidence with the K--S test that low-mass BDs are similar to the planetary population (and thus presumably formed in the disc), and that high-mass BDs belong to the stellar population (i.e. formed by GI of a molecular cloud akin to stellar formation). It is important to note however that the low-mass BD preference for metal-rich hosts may be biased by a potential tendency for metal-poor stars to be avoided in RV surveys, where most of the target stars in their study were originally identified \citep{Xiao2023}.  

\begin{figure*}
    \centering
    \includegraphics[width=0.90\linewidth]{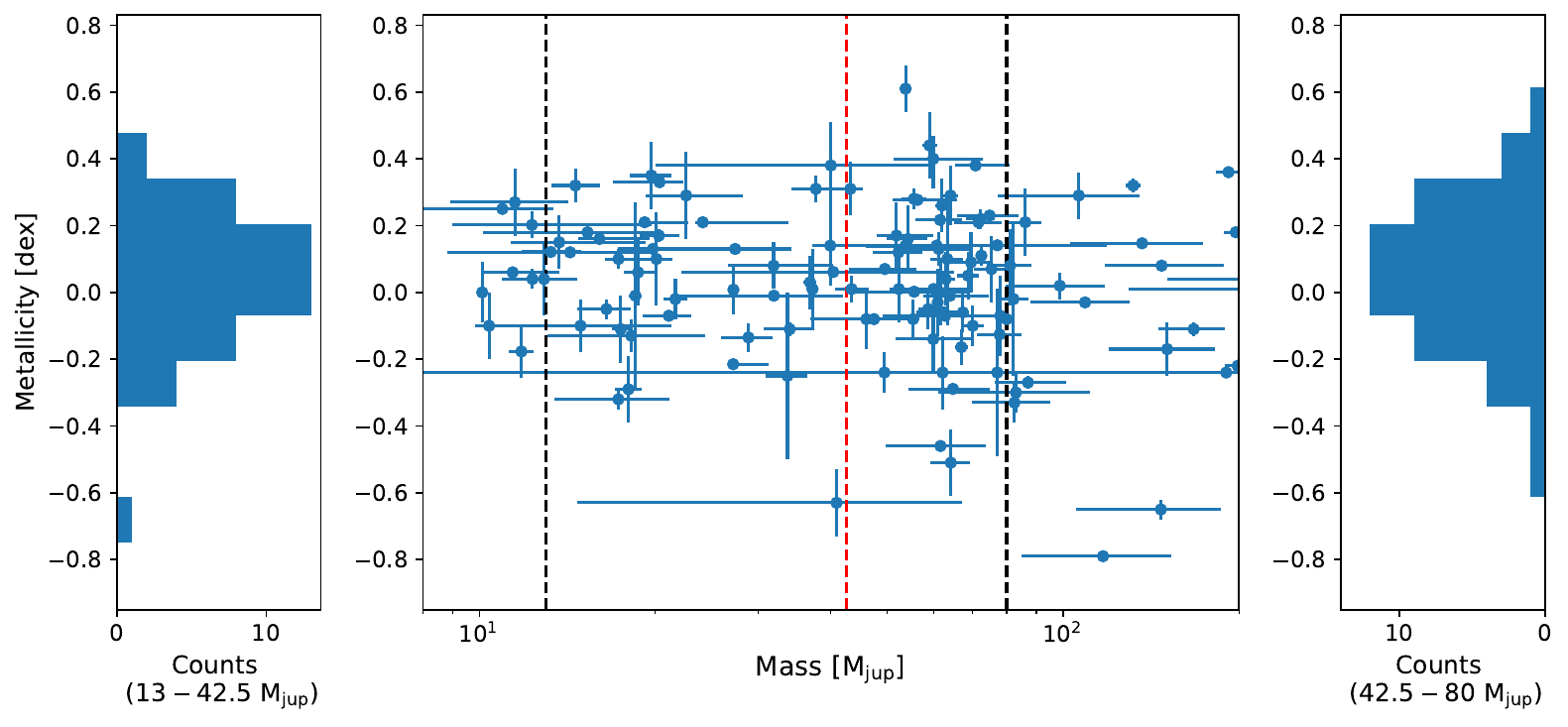}
    \caption{Host star metallicity plotted against companion mass, for objects with available measurement of metallicity from literature/SIMBAD. Histograms at either side of the main plot show the distributions of metallicity for low-mass (left) and high-mass BDs (right). The black vertical dashed lines indicate the BD mass regime, and the red dashed line denotes the $42.5$~M$_{\textrm{Jup}}$ transition \citep{MaGe2014}.}
    \label{fig:FeH_M}
\end{figure*}

To compare the metallicity distribution of stars in our sample, we have chosen to include all stars in our sample with metallicity information (in previous BD desert literature or on SIMBAD) to see if the same trends are observed. In Fig.~\ref{fig:FeH_M} we have plotted the BD region (bounded by the black dashed lines) as well as companions in the sample that have recently been re-classified as low-mass stars. 

Visual inspection of Fig.~\ref{fig:FeH_M} reveals that the low-mass and high-mass BDs do not show very distinct distributions, with histograms peaking at similar metallicities (albeit with high-mass BDs having a slightly wider distribution).
The BDs that have moved to VLMS regime appear to fit with trends for stars, where they are spread over a greater range and generally have sub-solar metallicity. Performing the two-sample K--S test on objects within ranges $13<M<42.5$~M$_{\textrm{jup}}$ or $42.5<M<80$~M$_{\textrm{jup}}$, we again do not find any evidence that the two populations are not drawn from the same parent distribution. When following the same criteria as \citet{Xiao2023} and selecting hosts more massive than $0.52$~M$_{\sun}$, we do not observe any statistical reason to reject the null hypothesis of a single overarching metallicity sample.

\section{Use of Gaia products}\label{sec:limitations}
There are a few caveats, or limitations, to using the NSS results in the manner that we have. We elaborate below. 

\subsection{Mapping minimum mass to BD mass}
It is apparent from Section~\ref{sec:BDverification} that in some cases the RV-derived minimum mass $M\sin{i}$ does not map directly to \texttt{binary\_masses} $M$ for a given inclination value. This is also noted by \citet{GaiaStellarMulti2023}, who find that \texttt{binary\_masses} values can occasionally be far larger, or smaller, than one would expect. 

\citet{GaiaStellarMulti2023} go on to discuss that when taking into account uncertainties on masses (that can be very large in \texttt{binary\_masses}), the discrepancies between either solving Kepler's third law with RV and inclination or calculating mass from the astrometric orbital fit directly are not very statistically significant, but nonetheless still need to be understood. 

Some disagreement could arise from differing values adopted for the mass of the primary. This will be a key factor in calculating the mass of the secondary, in either RV work or astrometric observations. The isochrone fitting performed to generate the \texttt{binary\_masses} (see Section~\ref{sub:true-masses} and \citealt{GaiaStellarMulti2023}) will differ from most of the individual methods used throughout RV studies of brown dwarfs and their host stars.

There are also other more fundamental effects that could cause the discrepancies, such as the sensitivity of \textit{Gaia} DR3, and selection effects and biases involved in the processing and creation of the NSS \citep{GaiaStellarMulti2023}.

\subsection{Orbital parameter mismatch}
Differences between the NSS solution and other astrometric orbital fits were recently observed by \citet{Xiao2023}. They used the cross-calibrated \textit{Hipparcos}--\textit{Gaia} catalogue of accelerations (HGCA) and the \textsc{orvara} fitting tool to determine masses for substellar objects. For orbits with period less than twice the baseline of Gaia ($\sim6$~yr), \citet{Xiao2023} remark that validation of their results compared to DR3 solutions is required. For the 65 stars in their sample that fall into this category, 12 have solutions amenable to inclination calculations in the NSS. 8 of the 12 are consistent with the NSS to within $1\sigma$, whereas 4 are not. This is as expected statistically, as a third of points lie outside of $1\sigma$. The masses from \textsc{orvara} are not consistently smaller or larger, but vary from case to case. 

Future data releases from \textit{Gaia} may help shed some light on the nature of these discrepancies by observing how solutions change with additional data, both through the NSS reduction pipeline and via accelerations and \textsc{orvara}. DR4 is expected to be published after the end of 2025, and the catalogue will contain all available non-single star solutions for 66 months of astrometric, spectroscopic, and photometric data.

\section{Conclusions}\label{sec:conclusions}
We have combed the \textit{Gaia} DR3 NSS archive and the literature to perform mass determinations of BD candidates. Our results in summary are:

\begin{enumerate}
    \item We have created an up-to-date catalogue of 214 BDs in binaries with $P < 10^{4}$\,d. This allows us to examine the population statistics of the brown dwarf desert.
    \item Using this new catalogue, we have searched \textit{Gaia} DR3 NSS results for inclinations of BD orbits allowing companion masses to be calculated from minimum masses. 12 BD candidates were updated in this way, with 3 remaining desert BDs and 9 moving to the stellar regime.
    \item Among these 12 candidates, we have been able to break degeneracies in or find solutions for archival RVs for three previously known BD candidates.
    \item We identified a further 19 BD candidates with mass in the BD range and with periods less than $\sim 1200$\,d in the DR3 \texttt{binary\_masses} database.
    \item Our results corroborate findings of a valley in the mass distribution with minimum around $30$--$35$~M$_{\textrm{jup}}$, and that periods $<100$~d are still under-populated in comparison to longer periods
    \item Masses derived by combining $M\sin{i}$ with the NSS inclination do not always map directly to the purely astrometric secondary masses/mass bounds in the \texttt{binary\_masses} database. Of the 12 derived masses in Section~\ref{sec:BDverification}, ten of these have astrometric masses, and 4 of these are not consistent with the solved RV minimum mass to within $2\sigma$ of the error on the derived masses. Further work is required to determine why this is the case.
    \item When examining the mass-eccentricity distribution, a split into two groups of high and low-mass objects is only marginally statistically significant, quantified by a two--sided K--S test $p$-value of 0.09~per~cent. This $p$-value becomes 0.05~per~cent, just equal to the threshold for rejecting the null hypothesis, when taking a split in mass at $40$~M$_{\textrm{jup}}$. This result hints at two different parent distributions, and two potential origins -- either akin to planetary formation, or stellar.
    \item We identify that the period-eccentricity plane shows no low eccentricity objects at periods around $100$~d.
    \item We find no evidence of low and high-mass BDs being split by metallicity, indicating that core accretion is not the dominant formation mechanism for BDs as they do not follow the same trends that giant exoplanets do with metallicity.
    \item By plotting all BDs that are either updated by DR3 information, or are new candidates in the release, we identify a diagonal envelope bounding the \textit{Gaia} points, highlighting the sensitivity of the currently available NSS solutions from 34 months of data.
    \item The limitations of this work have been discussed, and it is worth re-iterating them for others wishing to use the NSS results for similar purposes. This study also shows the positives of using DR3 results to investigate stellar companions, and how these can quickly be applied to characterise BDs.
\end{enumerate}

\section*{Acknowledgements}


ATS is supported by a Science and Technology Facilities Council (STFC) studentship. CAH and JRB are supported by grants ST/T000295/1 and ST/X001164/1 from STFC. JKB is supported by an STFC Ernest Rutherford Fellowship (grant ST/T004479/1). This research has made use of the SIMBAD data base, operated at CDS, Strasbourg, France. This research has made use of data obtained from the portal \href{http://exoplanet.eu}{\texttt{exoplanet.eu}} of The Extrasolar Planets Encyclopaedia. This research has made use of the Washington Double Star Catalog maintained at the U.S. Naval Observatory.

Radial velocity data were analysed with the \textsc{exo-striker} software (\url{https://github.com/3fon3fonov/exostriker}). 

This work has made use of data from the European Space Agency (ESA) mission {\it Gaia} (\url{https://www.cosmos.esa.int/gaia}), processed by the {\it Gaia} Data Processing and Analysis Consortium (DPAC, \url{https://www.cosmos.esa.int/web/gaia/dpac/consortium}). Funding for the DPAC has been provided by national institutions, in particular the institutions participating in the {\it Gaia} Multilateral Agreement.

Funding for SDSS-III has been provided by the Alfred P. Sloan Foundation, the Participating Institutions, the National Science Foundation, and the U.S. Department of Energy Office of Science. The SDSS-III web site is \url{http://www.sdss3.org/}.

SDSS-III is managed by the Astrophysical Research Consortium for the Participating Institutions of the SDSS-III Collaboration including the University of Arizona, the Brazilian Participation Group, Brookhaven National Laboratory, Carnegie Mellon University, University of Florida, the French Participation Group, the German Participation Group, Harvard University, the Instituto de Astrofisica de Canarias, the Michigan State/Notre Dame/JINA Participation Group, Johns Hopkins University, Lawrence Berkeley National Laboratory, Max Planck Institute for Astrophysics, Max Planck Institute for Extraterrestrial Physics, New Mexico State University, New York University, Ohio State University, Pennsylvania State University, University of Portsmouth, Princeton University, the Spanish Participation Group, University of Tokyo, University of Utah, Vanderbilt University, University of Virginia, University of Washington, and Yale University. 

Based on data retrieved from the SOPHIE archive at Observatoire de Haute-Provence (OHP), available at \href{http://atlas.obs-hp.fr/sophie}{\texttt{atlas.obs-hp.fr/sophie}}

Plots in this paper have used color maps specifically designed to be colorblind friendly, created by \citet{Wong2011}.

The following python modules have been used in this work: \textsc{numpy}, \textsc{scipy}, \textsc{astropy}, \textsc{emcee}, \textsc{matplotlib}, \textsc{ndtest}, \textsc{nsstools}.

\section*{Data Availability}

The created catalogue underlying this article is available in Open Research Data Online (ORDO; \url{https://ordo.open.ac.uk/}) at \url{https://doi.org/10.21954/ou.rd.24156393} as well as on github at \url{https://github.com/adam-stevenson/brown-dwarf-desert/tree/main}. The archival radial velocities used are all readily available at SOPHIE and SDSS archives.



\bibliographystyle{mnras}
\bibliography{refs} 




\appendix

\section{Orbital Significances}
As part of the NSS pipeline, astrometric solutions are provided where the significance ($s$) of the orbit is $s>5$. The significance\footnote{\url{https://gea.esac.esa.int/archive/documentation/GDR3/Data_analysis/chap_cu4nss/sec_cu4nss_astrobin/ssec_cu4nss_astrobin_orbital.html}} is calculated through dividing the semi-major axis of the orbit by its uncertainty,

\begin{equation}\label{eqn:NSS_significance}
    s = \frac{a_{0}}{\sigma_{a_{0}}}.
\end{equation}

\noindent Orbital solution searches begin with a period search, that may lead to detection of periods relating to the \textit{Gaia} scanning law \citep{GaiaStellarMulti2023}. Many periods less than 100~d were found to be erroneous, and the solutions needed to be removed. An additional significance requirement was added, of $s >\frac{158}{P_{\textrm{days}}}$ \citep{Halbwachs2023}.

To confirm that all companion orbits discussed in this paper are indeed of good significance, we have listed these values in Table~\ref{tab:orbitSignificance}. Both $a_{0}$ and $\sigma_{a_{0}}$ were calculated with \textsc{nsstools} from the Thiele-Innes elements $A$, $B$, $F$ and $G$.

\renewcommand{\arraystretch}{1.00}
\begin{table}
	\centering
	\caption{Orbital signficances of astrometric orbits in the NSS, for all companions in this work. Calculated with equation~\ref{eqn:NSS_significance}, all are greater than $5$, showing that semi major axis is known to better than 20 per cent. Listed in order corresponding to Tables~\ref{tab:NewTrueMasses} and \ref{tab:BD_bin_mass}.}
	\label{tab:orbitSignificance}
	\hspace*{-0.9cm}\begin{tabular}{lc} 
		\hline
		Host Star & Significance ($a_{0}/ \sigma_{a_{0}}$) \\
		\hline
        HD\,39392 & $5.24$\\
		  HD\,132032 & $14.88$ \\
		HD\,140913 & $25.68$ \\
        HIP\,67526 & $23.88$ \\
		BD+24\,4697 & $87.40$ \\
        BD+26\,1888 & $84.16$ \\
        HD\,30339 & $10.51$  \\
        HD\,105963\,A & $616.42$  \\
        HD\,160508 & $23.91$ \\
        TYC~0173-02410-1 & $59.88$ \\
        GSC~03467-00030 & $33.88$  \\
        2MASS J04422788+0043376 & $35.21$ \\
        G 165-52 & $46.71$ \\
        GSC 04516-00523 & $18.28$  \\
        HD\,104289 & $6.48$  \\
        HD\,115517 & $19.06$  \\
        HD\,156312\,B & $27.01$  \\
        HIP\,117179 & $13.70$  \\
        HIP\,60321 & $84.34$  \\
        HIP\,75202 & $27.77$  \\
        LP 498-48 & $87.10$  \\
        LSPM J1657+2448 & $32.76$  \\
        LSPM J1831+4213 & $12.05$  \\
        TYC 3056-264-1 & $45.14$  \\
        TYC 3873-761-1 & $33.80$  \\
        TYC 7572-327-1 & $17.26$  \\
        TYC 8321-266-1 & $16.58$  \\
        TYC 9255-929-1 & $27.34$  \\
        UCAC2 9182345 & $26.24$  \\
        UCAC4 302-050985 & $50.39$ \\
		\hline
	\end{tabular}
\end{table}

\bsp	
\label{lastpage}
\end{document}